\documentclass[twocolumn]{aastex631}

\usepackage{amsmath,amstext}
\usepackage[T1]{fontenc}
\usepackage{apjfonts}
\usepackage{multirow}
\usepackage{xcolor}

\newcommand{\hst}{\textit{HST}}

\newcommand{\jwst}{\textit{JWST}}

\newcommand{\hb}{\hbox{H$\beta$}}
\newcommand{\ha}{\hbox{H$\alpha$}}
\newcommand{\oii}{\hbox{[O \textsc{ii}]}}
\newcommand{\oiii}{\hbox{[O \textsc{iii}]}}
\newcommand{\nii}{\hbox{[N \textsc{ii}]}}
\newcommand{\galfit}{\hbox{\textsc{galfit}}}
\newcommand{\cigale}{\hbox{\textsc{cigale}}}
\newcommand{\linmix}{\hbox{\textsc{linmix}}}
\newcommand{\grizli}{\hbox{\textsc{grizli}}}
\newcommand{\bagpipes}{\hbox{\textsc{bagpipes}}}
\newcommand{\statmorph}{\hbox{\textsc{statmorph}}}

\begin{document}

\title{NGDEEP Epoch 1: Spatially Resolved \ha\ Observations of Disk and Bulge Growth in Star-Forming Galaxies at z~$\mathbf{\sim}$~0.6--2.2 from JWST NIRISS Slitless Spectroscopy}

\correspondingauthor{Lu Shen}
\email{lushen@tamu.edu}

\author[0000-0001-9495-7759]{Lu Shen}
\affiliation{Department of Physics and Astronomy, Texas A\&M University, College Station, TX, 77843-4242 USA}
\affiliation{George P.\ and Cynthia Woods Mitchell Institute for
 Fundamental Physics and Astronomy, Texas A\&M University, College Station, TX, 77843-4242 USA}
 
\author[0000-0001-7503-8482]{Casey Papovich}
\affiliation{Department of Physics and Astronomy, Texas A\&M University, College Station, TX, 77843-4242 USA}
\affiliation{George P.\ and Cynthia Woods Mitchell Institute for
 Fundamental Physics and Astronomy, Texas A\&M University, College Station, TX, 77843-4242 USA}

\author[0000-0002-7547-3385]{Jasleen Matharu}
\affiliation{Cosmic Dawn Center (DAWN), Denmark}
\affiliation{Niels Bohr Institute, University of Copenhagen, Jagtvej 128, DK-2200 Copenhagen N, Denmark}

\author[0000-0003-3382-5941]{Nor Pirzkal}
\affiliation{ESA/AURA Space Telescope Science Institute}

\author[0000-0003-3424-3230]{Weida Hu}
\affiliation{Department of Physics and Astronomy, Texas A\&M University, College Station, TX, 77843-4242 USA}
\affiliation{George P.\ and Cynthia Woods Mitchell Institute for
 Fundamental Physics and Astronomy, Texas A\&M University, College Station, TX, 77843-4242 USA}

 \author[0000-0001-8534-7502]{Bren E. Backhaus}
\affiliation{Department of Physics, 196 Auditorium Road, Unit 3046, University of Connecticut, Storrs, CT 06269}

\author[0000-0002-9921-9218]{Micaela B. Bagley}
\affiliation{Department of Astronomy, The University of Texas at Austin, Austin, TX, USA}

\author[0000-0001-8551-071X]{Yingjie Cheng}
\affiliation{University of Massachusetts Amherst, 710 North Pleasant Street, Amherst, MA 01003-9305, USA}

\author[0000-0001-7151-009X]{Nikko J. Cleri}
\affiliation{Department of Physics and Astronomy, Texas A\&M University, College Station, TX, 77843-4242 USA}
\affiliation{George P.\ and Cynthia Woods Mitchell Institute for Fundamental Physics and Astronomy, Texas A\&M University, College Station, TX, 77843-4242 USA}

\author[0000-0001-8519-1130]{Steven L. Finkelstein}
\affiliation{Department of Astronomy, The University of Texas at Austin, Austin, TX, USA}

\author[0000-0002-1416-8483]{Marc Huertas-Company}
\affil{Instituto de Astrof\'isica de Canarias, La Laguna, Tenerife, Spain}
\affil{Universidad de la Laguna, La Laguna, Tenerife, Spain}
\affil{Universit\'e Paris-Cit\'e, LERMA - Observatoire de Paris, PSL, Paris, France}

\author[0000-0002-7831-8751]{Mauro Giavalisco}
\affiliation{University of Massachusetts Amherst, 710 North Pleasant Street, Amherst, MA 01003-9305, USA}

\author[0000-0001-9440-8872]{Norman A. Grogin}
\affiliation{Space Telescope Science Institute, Baltimore, MD, USA}

\author[0000-0003-1187-4240]{Intae Jung}
\affiliation{Space Telescope Science Institute, Baltimore, MD, 21218, USA}

\author[0000-0001-9187-3605]{Jeyhan S. Kartaltepe}
\affiliation{Laboratory for Multiwavelength Astrophysics, School of Physics and Astronomy, Rochester Institute of Technology, 84 Lomb Memorial Drive, Rochester, NY 14623, USA}

\author[0000-0002-6610-2048]{Anton M. Koekemoer}
\affiliation{Space Telescope Science Institute, 3700 San Martin Dr., 
Baltimore, MD 21218, USA}

\author[0000-0003-3130-5643]{Jennifer M. Lotz}
\affiliation{Gemini Observatory/NSF's National Optical-Infrared Astronomy Research Laboratory, 950 N. Cherry Ave., Tucson, AZ 85719, USA}

\author[0000-0003-0695-4414]{Michael V. Maseda}
\affiliation{Department of Astronomy, University of Wisconsin-Madison, 475 N. Charter St., Madison, WI 53706 USA} 

\author[0000-0003-4528-5639]{Pablo G. P\'erez-Gonz\'alez}
\affiliation{Centro de Astrobiolog\'{\i}a (CAB), CSIC-INTA, Ctra. de Ajalvir km 4, Torrej\'on de Ardoz, E-28850, Madrid, Spain}

\author[0000-0003-2283-2185]{Barry Rothberg}
\affiliation{U.S. Naval Observatory, 3450 Massachusetts Avenue NW, Washington, DC 20392, USA}
\affiliation{Department of Physics and Astronomy, George Mason University, 4400 University Drive, MSN 3F3, Fairfax, VA 22030, USA}

\author[0000-0002-6386-7299]{Raymond C. Simons}
\affiliation{Department of Physics, 196 Auditorium Road, Unit 3046, University of Connecticut, Storrs, CT 06269}

 \author[0000-0002-8224-4505]{Sandro Tacchella}
\affiliation{Kavli Institute for Cosmology, University of Cambridge, Madingley Road, Cambridge, CB3 0HA, UK} 
\affiliation{Cavendish Laboratory, University of Cambridge, 19 JJ Thomson Avenue, Cambridge, CB3 0HE, UK}

\author[0000-0003-2919-7495]{Christina C. Williams}
\affiliation{NSF's National Optical-Infrared Astronomy Research Laboratory, 950 N. Cherry Avenue, Tucson, AZ 85719, USA} 
\affiliation{Steward Observatory, University of Arizona, 933 North Cherry Avenue, Tucson, AZ, 85721, USA}

\author[0000-0003-3466-035X]{{L. Y. Aaron} {Yung}}
\affiliation{Astrophysics Science Division, NASA Goddard Space Flight Center, 8800 Greenbelt Rd, Greenbelt, MD 20771, USA}



\begin{abstract}


We study the \ha\ equivalent width, EW(\ha), maps of 19 galaxies at $0.6 < z < 2.2$ in the Hubble Ultra Deep Field (HUDF) using NIRISS slitless spectroscopy as part of the Next Generation Deep Extragalactic Exploratory Public (NGDEEP) Survey.  
Our galaxies mostly lie on the star-formation main sequence with stellar masses between $\mathrm{10^9 - 10^{11}\ M_\odot}$, characterized as ``typical'' star-forming galaxies at these redshifts. 
Leveraging deep \hst\ and \jwst\ images, spanning 0.4--4.8~$\mu$m, we perform spatially-resolved fitting of the spectral energy distributions (SEDs) for these galaxies and construct specific star formation rate (sSFR) and stellar-mass-weighted age maps with a spatial resolution of $\sim$1 kpc.  
The pixel-to-pixel EW(\ha) increases with increasing sSFR and with decreasing age. The average trends are slightly different from the relations derived from integrated fluxes of galaxies from the literature, suggesting complex evolutionary trends within galaxies. 
We quantify the radial profiles of EW(\ha), sSFR, and age. The majority (84\%) of galaxies show positive EW(\ha) gradients, in line with the inside-out quenching scenario.  
A few galaxies (16\%) show inverse (and flat) EW(\ha) gradients possibly due to merging or starbursts. 
We compare the distributions of EW(\ha) and sSFR to star formation history models (SFHs) as a function of galactocentric radius. We argue that the central regions of galaxies have experienced, at least one, rapid star-formation episodes, which leads to the formation of the bulge, while their outer regions (e.g., disks) grow via more smoothly varying SFHs.   
These results demonstrate the ability to study resolved star formation in distant galaxies with \jwst\ NIRISS.

\end{abstract}


\section{Introduction} \label{sec:intro}

%
Investigating spatially resolved properties of galaxies offers important constraints on the mechanisms by which galaxies form stars and grow in stellar mass (e.g., \citealp{Brinchmann2004,Papovich2005}).
Studies on the structure of galaxies up to $z \sim 2.5$ have converged to a coherent picture that star-forming galaxies (SFGs) are on average larger and disk-dominated systems, while quiescent galaxies are more compact and bulge-dominated (e.g., \citealp{Shen2003, Wuyts2011, Weinzirl2011, vanderWel2014, vanDokkum2015}). 
These studies seem to imply that as SFGs evolve and quench their star formation, they may experience bulge assembly and/or compaction \citep{Dekel2014, Papovich2015,Zolotov2015, Tacchella2016b}, which results in an inside-out growth/quenching pattern that the central regions are formed at earlier times and therefore show older age and lower specific star-formation rates ($\mathrm{sSFR\equiv SFR / M_*}$) than the outer disc. 
This inside-out quenching has been supported by extensive observational evidence from spatially-resolved studies at low redshift that the sSFR profiles appear to be centrally suppressed in massive galaxies $\mathrm{log(M_*/M_\odot)} > 10.6$ \citep[e.g.][]{Spindler2018}, in green valley galaxies \citep[e.g.][]{Belfiore2018}, and in galaxies below the star formation main sequence (SFMS; e.g., \citealp{Ellison2018}).

At higher redshift ($z \gtrsim 1$), studies have used \ha\ to measure spatially-resolved star formation in galaxies using the integral field units (IFU; \citealp{Tacchella2015, Tacchella2018, Wilman2020}) and the \hst\ Grisms \citep{Nelson2012, Nelson2016b, Matharu2022}. These studies have found evidence that ongoing star formation occurs in disks that are more extended than those occupied by existing stars in SFGs at $z\sim0.5 - 2.7$, and the interpretation is that the disks form ``inside-out''.
\citet{Tacchella2018} focused on ten massive galaxies at $z\sim2$ and found flat sSFR radial profiles for galaxies with $\mathrm{M_* = 10^{10.5 - 11} M_\odot}$, and found suppressed central sSFR radial profiles for galaxies with $\mathrm{M_* > 10^{11} M_\odot}$.  They interpreted this as galaxies being quenched from the inside out. 
This scenario is also supported by the study of the gas reservoirs in a massive star-formation galaxy at $z\sim 2.2$ that shows that both molecular gas fraction and molecular gas depletion time are lower in the central region compared to the outskirts  \citep{Spilker2019}. 
However, these studies are limited to massive and brighter galaxies or are limited to lower redshifts because \hst\ can only study \ha\ to $z < 1.6$, or rely on ``stacking'' to get a sufficient signal-to-noise ratio.


Recently, there has been an increasing number of studies focusing on the 2D and radial properties of high-redshift galaxies using \jwst\ data \citep{Perez-Gonzalez2023a, Gimenez-Arteaga2023, Wang2022, Abdurrouf2023}. 
In this paper, we build on these using measurements of the spatially resolved \ha\ derived from slitless grism spectroscopy from \jwst\ \citep{Gardner2006, Gardner2023} NIRISS \citep{Rene2023} for 19 galaxies at $0.6 < z < 2.2$, using the first epoch of the deep \jwst\ NIRISS slitless spectroscopy from the Next Generation Deep Extragalactic Exploratory Public (NGDEEP) Survey \citep{Bagley2022}. 
We focus on the equivalent width (EW) of \ha\ which provides an independent method for determining the sSFR, as the EW(\ha) is a ratio of \ha\ flux, an SFR indicator, and the continuum flux at $\lambda_{rest}=6563$ \AA\ which traces the stellar mass. Additionally, EW(\ha) is less susceptible to the presence of dust compared to the rest-frame UV. 
As a comparison, we explore the sSFR and stellar mass-weighted age maps of these galaxies derived from a spatially resolved SED fitting method on similar spatial resolution using 9 bands from \textit{Hubble Space Telescope} (\hst) ACS and WFC3 imaging from CANDELS, and 11 bands from \jwst\ imaging from NGDEEP, the JWST Advanced Deep Extragalactic Survey (JADES, \citealp{Eisenstein2023, Rieke2023}) and the JWST Extragalactic Medium-band Survey (JEMS, \citealp{Williams2023}).  
We compare these spatially resolved properties and their radial profiles to reveal the spatial distribution of stellar mass, star formation, and age of these galaxies.  
Thanks to the high spatial resolution, deep imaging, and slitless spectroscopy, we can perform these spatially-resolved analyses on individual galaxies, which provide details on the star formation, such as those contributed from off-center clumps that are mostly erased when stacking. 


Throughout this paper, all magnitudes, are presented in the AB system \citep{Oke1983, Fukugita1996}. We adopt a standard $\Lambda$--cold dark matter ($\Lambda$CDM) cosmology with $H_0$ = 70 km s$^{-1}$, $\Omega_{\rm \Lambda,0}$ = 0.70, and $\Omega_{\rm M,0}$ = 0.30 \citep{Planck2016}.

\section{Data and Sample Selection } \label{sec:data}

\subsection{The NGDEEP survey} \label{sec:survey}

We use the \jwst\ NIRISS Wide Field Slitless Spectroscopy (WFSS) observations of galaxies as part of NGDEEP (Proposal ID \#2079; \citealp{Bagley2023}). 
NGDEEP targets the Hubble Ultra Deep Field (HUDF) with NIRISS slitless spectroscopy. 
The science aims of NGDEEP include measuring metallicities and SFRs for low-mass galaxies in the redshift range $0.5 < z < 4$. 
In parallel, NGDEEP targets the HUDF-Par2 parallel field with NIRCam with the science aim of discovering galaxies up to $z > 12$ and constraining the slope of the faint-end of the rest-UV luminosity function \citep{Leung2023}. 

The first epoch of NGDEEP observations was taken over 2023 Jan 31 - Feb 2.  These include observations with NIRISS with the GR150R and GR150C grisms and the F115W, F150W, and F200W filters. 
The total exposure times on grisms are 95 ks in F115W, 43 ks in F150W, and 32 ks in F200W. In addition, the observations include NIRISS direct imaging in the same filters with total integration times of 5.4, 1.7, and 1.7 ks, respectively. 

We reduced the direct images using {v1.9.4} of the \jwst\ Pipeline\footnote{\url{https://github.com/spacetelescope/jwst}} \citep{Bushouse2023}.  
From the direct images with NIRISS, we measured the 3$\sigma$ depth of F115W, F150W, and F200W imaging with $0\farcs2$ radius aperture to be 29.1, 28.6, and 28.6 AB mag, respectively. 
For the grism spectroscopic imaging, we reduced the NIRISS data using two methods, including the Grism Redshift \& Line software (\grizli; \citealp{Brammer2022}) and EMission-line two-Dimensional (\textsc{ew2d}, \citealp{Pirzkal2018}). The latter method is described in \citet{Pirzkal2023}. For the analysis of emission-line maps here, we used the images from \grizli\ (see details in Section \ref{sec:grism} below). 
From the one-dimensional extracted spectra, we found that the grism data are sensitive to line fluxes to $f_\mathrm{lim}\approx 2 \times10^{-18}$ erg~s$^{-1}$ cm$^{-2}$ (3$\sigma$, \citealt{Pirzkal2023}). 


\subsection{Available Optical/NIR Imaging and Photometry} \label{sec:imaging}

We utilized a vast array of deep imaging taken with \hst\ and \jwst\ available in the HUDF field. 
This includes \hst\ ACS F435W, F606W, F775W, F814W, F850LP, and WFC3/IR F105W, F125W, F140W and F160W, and we use the latest reductions processed as part of the Cosmic Assembly Near-IR Deep Extragalactic Legacy Survey (CANDELS, \citealp{Koekemoer2011, Grogin2011}), see \citet{Finkelstein2022} for more details. 
Besides the three \jwst\ NIRISS direct-imaging bands from NGDEEP, we include the nine \jwst\ NIRCam bands: F090W,  F335M, F356W, and F444W from the DR1 release from JADES \citep{Eisenstein2023, Rieke2023} and F182M, F210M, F430M, F460M and F480M from the DR1 release from JEMS \citep{Williams2023}. 
%
%
%

We adopt the \hst\ photometry catalog from \citet{Finkelstein2022}, and use the same method to extract the photometry in \jwst\ imaging. 
To measure consistent photometry across all bands, we align each image astrometrically and on the same pixel scale ($0\farcs06$). We then match each image to the PSF of the F160W image. 
{The PSFs of HST images are constructed by stacking selected stars (see \citealt{Finkelstein2022} for details). 
However, there are no isolated, unsaturated stars available in the NGDEEP NIRISS field with which to construct an empirical PSF.
Thus,} we constructed a PSF for each \jwst\ bandpass using WebbPSF \citep{Perrin2012, Perrin2014}. {For the PSF in each bandpass, we generated 100 different PSFs at different position angles, which we then averaged.}

%

We constructed kernels to match each PSF to that in the \hst\ WFC3/IR F160W band using the pypher\footnote{https://pypher.readthedocs.io} Python routine. Each image was then convolved with its respective kernel. {To examine the accuracy of this PSF matching process, we measured the growth curve of four, compact elliptical galaxies in the PSF-matched images. The average ratio of the growth curve of these elliptical galaxies in the PSF-matched JWST images and the HST F160W image spans a range of 0.99 to 1.10 at a radius of 1\arcsec. This was one of the motivating factors to include a 10\% additional error (in quadrature) to the uncertainties of the fluxes when we performed the integrated and spatial resolved SED fitting. }

We use the F160W image as the detection image and cycle through every aligned and PSF-matched \jwst\ image as the measurement. 
We adopt the small Kron elliptical apertures and correct to total using an aperture correction. We estimate photometry uncertainties empirically to account for the Poisson correlation following \citet{Finkelstein2022}.


\subsection{Grism Data Reduction and Spectral Extractions} \label{sec:grism}

For the analysis of emission-line maps here, we processed the data using the latest version of the \grizli.
\grizli\ performs full end-to-end processing of NIRISS imaging and slitless spectroscopic data sets, including retrieving, pre-processing the raw observations for cosmic rays, flat-fielding, sky subtraction, astrometric corrections, alignment, modeling contamination from overlapping spectra, extracting 1D and 2D spectra, and fitting full continuum+emission-line models.  

From the data products, \grizli\ derives the emission-line maps by drizzling the contamination- and continuum-subtracted 2D spectral beams back to the imaging plane.  It uses the \jwst/NIRISS imaging to scale the spectroscopic data to match the total fluxes measured in the direct images. 
We created emission-line maps with a pixel scale of $0\farcs1$, similar to the FWHM of PSF in our NIRISS filters. The uncertainties on the line maps are calculated using the drizzle weights from the constituent beam pixels. 
For additional details on \grizli\ and its data products, we refer the reader to \cite{Estrada-Carpenter2019, Simons2021, Matharu2021, Papovich2022,Wang2022, Simons2023, Noirot2023, Matharu2023}. 

\begin{figure*}
    \centering
    \includegraphics[width=\textwidth]{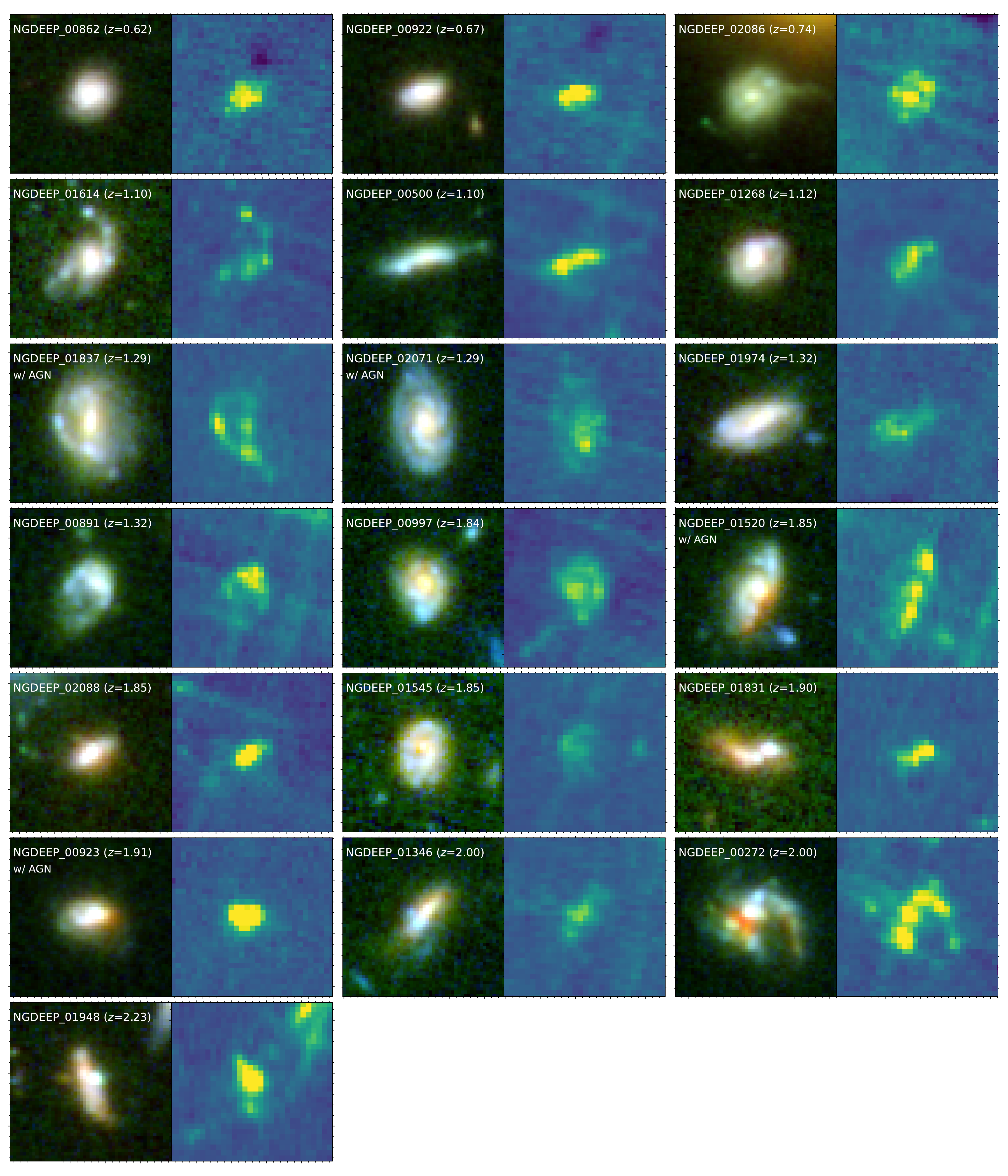}
    \caption{False color images and H$\alpha$ maps for the galaxies in our galaxy sample.  For each galaxy, we show the RGB images using the \hst/ACS F814W, \jwst/NIRISS F150W and \jwst/NIRCam F444W (\textit{left}), and the \ha\ emission line map constructed from the NIRISS slitless spectroscopy (\textit{right}).  The sample is ordered by increasing redshift. Each image is $2^{\prime\prime}\times2^{\prime\prime}$, corresponding to 13.4$\times$13.4 kpc${^2}$ and 16.6$\times$16.6 kpc${^2}$ at $z=0.6$ and $z=2.2$, respectively. The text inset lists the galaxy ID, redshift, and whether galaxies are identified with AGN. }
    \label{fig:stamps}
\end{figure*}

\subsection{Sample Selection} \label{sec:sample}

Here, our goal is to study the emission-line maps for galaxies with high signal-to-noise ratios (SNR). To construct this sample, we selected galaxies that have \ha\ line maps where the total \ha\ SNR is $>20\sigma$, and the galaxies have data in at least 50 pixels each with \ha\ SNR $>3\sigma$. The latter criterion is to ensure that we have enough pixels to derive the EW(\ha) radial gradient. For example, a circular area with $\ge$50 pixels would have a radius of $\ge$4 pixels, providing at least two annular bins each with a width of at least two pixels to facilitate a measurement of the gradient. 
We visually examined these \ha\ line maps and excluded those with any contamination across the source possible due to the failure of the continuum subtraction. 

Our final sample includes a total of 19 galaxies. 
Fig.\ref{fig:stamps} displays the false color images and \ha\ line maps for this sample. 
We note that the current NIRISS grism data has 2 position angles (GR150R and GR150C grisms), providing good spatial coverage of the emission-line features.  The full NGDEEP NIRISS data will double the data and have four position angles, which should largely reduce the fraction of contamination and increase the number of galaxies with secure line maps. 

The \grizli-derived redshifts are consistent with previous ground-based and space-based spectroscopic measurements of our sources, where we find a median offset of only $\Delta z/(1 + z)$ = 0.002. Therefore we are confident in the identification of the \ha\ emission line for these galaxies.

We identify four galaxies in our sample as AGN by cross-matching to the latest AGN catalog from \citet{Lyu2022}. 
Two of them (NGDEEP\_01837 and NGDEEP\_02071) are classified as AGN based on their near- to mid-IR excess emission from SED fitting and the other two (NGDEEP\_01520 and NGDEEP\_00923) are classified as AGN based on their X-ray to radio luminosity ratio higher than expected for stellar processes in a galaxy.  
For these galaxies, to remove any AGN contamination, we mask their central $3\times 3$ pixels ($0\farcs3 \times 0\farcs.3$, larger than the PSF FWHM of F160W image), where the center is derived from the peak of the F160W image. 

\begin{figure}
    \centering
    \includegraphics[width=\columnwidth]{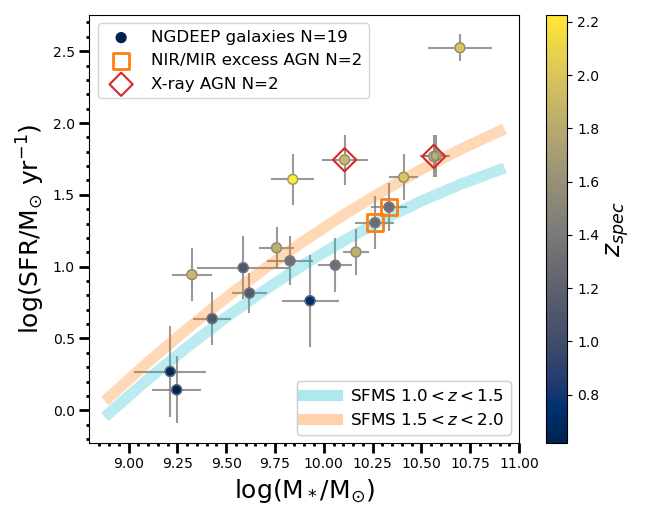}
    \caption{The stellar mass versus {the SED-derived SFR} of our final sample. Galaxies are colored by their redshift. The two galaxies with X-ray-detected AGN are marked by red diamonds, and the two with AGN implied by their near-/mid-IR excess are marked with orange squares. For comparison, the star-formation--main- sequence (SFMS) for galaxies at $1.0 < z < 1.5$ and $1.5 < z < 2.0$ from \citet{Whitaker2014} are shown in blue and orange, respectively, as labeled.}
    \label{fig:sample}
\end{figure}

Figure~\ref{fig:sample} shows the stellar mass versus SFR of our sample using quantities obtained from SED fitting to the integrated emission (see Section \ref{sec:sed}). We compared our galaxies to the ``star formation main sequence'' (SFMS) derived from larger samples from \citet{Whitaker2014} in the redshift range of our sample. Our sample generally follows the SFMS with a median of $\mathrm{log(SFR/SFR_{MS})} = 0.05$ with a scatter from -0.05 to 0.31 dex from the 16th/84th percentile. Therefore, our sample represents galaxies along the main sequence of star formation at these redshifts. 
Although we preferentially select galaxies with extended \ha, we do not expect our sample to be biased toward galaxies with more star formation, because no significant correlation is found between SFR and effective radius at fixed stellar mass for galaxies at $0.5 < z < 2.5$ and stellar mass within $9.0 \le \mathrm{log(M_*/M_\odot)} < 11.0 $ \citep{Lin2020}.

\section{Methodology} \label{sec:methods}

\subsection{SED fitting} \label{sec:sed}

We employed the SED fitting Code Investigating GALaxy Emission (\cigale) \citep{Boquien2019, Yang2020} to derive constraints on the integrated properties of the stellar populations of our galaxies using the {21 bands} photometry catalog {spanning the wavelength range of 0.43 - 4.8 $\mu$m (see details in Section \ref{sec:imaging})} and the spatially-resolved properties using the {21-band} fluxes in each Voronoi bin (see Section \ref{sec:resolvedmaps}). The parameters for these fits are listed in Table~\ref{tab:sed}. 
In detail, we adopted a delayed exponential star formation history (SFH) allowing the $\tau$ and stellar age to vary from 0.05--20~Gyr and 0.02--10~Gyr, respectively. 
We assumed a \citet{Chabrier2003} IMF and the stellar population synthesis models presented by \citet{Bruzual2003} with solar (Z$_\odot$) metallicity. 
The dust attenuation follows the extinction law of \citet{Calzetti2000} 
 {allowing the dust attenuation in emission lines from nebular regions $E(B-V)_l$ to vary from 0 to 1.1, and a lower dust attenuation in stellar continuum with a fixed dust attenuation ratio ($E(B-V)_\mathrm{factor} = 0.44$) between emission lines and stellar continuum. This corresponds to allowing dust attenuation in the stellar continuum to vary from 0 to 0.5. }
The amplitude of the absorption UV bump feature produced by dust at 2175 \AA\ is allowed to range from 0 to 3, and the slope of the dust power law is allowed to vary $-$0.5 to 0. 

For the integrated properties, e.g., the mass,  SFR, and age, we also test the robustness of these properties by including mid-- and far--IR data from IRAC/Spitzer, MIPS/Spitzer from the GOODS-Herschel catalog (\citealp{Elbaz2011}; see discussion in Section~\ref{sec:dust}). Therefore, to fit the dust emission, we also adopt dust emission templates from \citet{Dale2014}, which model the star-forming component as $dM_d(U) \propto U^{-\alpha} dU$, where $M_d$ is the dust mass, $U$ is the radiation field intensity. We allow the $\alpha$ to vary from 0.25 - 4. 

We adopt quantities and associated errors for each parameter from the posteriors from the CIGALE output catalog. 
In particular, we use the stellar mass ($\mathrm{M_*}$), SFRs averaged over 100~Myr ($\mathrm{SFR_{100Myr}}$), and the mass-weighted-stellar age, and their associated errors. The sSFR is calculated as $\mathrm{sSFR\equiv SFR_{100Myr} / M_*}$.


\begin{deluxetable}{c|c}
\tablecaption{Parameters used in the SED fitting with \textsc{cigale}. \label{tab:sed}}
\tablewidth{0pt}
\tablehead{
\colhead{Parameter} & \colhead{Values}}
\startdata
\multicolumn{2}{c}{Star formation history (sfhdelayed) }  \\
\hline
\multirow{2}{*}{$\tau$ [Myr]}  &  50, 100, 300, 500, 1000, \\
& 2000, 3000, 5000, 10000, 20000 \\
\multirow{2}{*}{Age [Myr]}  & 20, 40, 60, 80, 100, 150, 200, 250, 500,\\
& 1000, 2000, 3000, 5000, 7000, 10000\\
\hline
\multicolumn{2}{c}{Simple stellar population \citep{Bruzual2003}} \\
\hline
IMF & \citet{Chabrier2003} \\
Metallicity &  0.02 \\ 
\hline
\multicolumn{2}{c}{Dust Attenuation \citep{Calzetti2000}}  \\
\hline
\multirow{2}{*}{E(B-V)$_{l}$} & 0.005, 0.01, 0.025, 0.05, 0.1, 0.15,\\
& 0.2, 0.3, 0.5, 0.7, 0.9, 1.1 \\
E(B-V)$_{\rm factor}$ & 0.44 \\
Amplitude of the UV bump & 0, 1.5, 3.0 \\
Slope of the power law & -0.5, -0.4, -0.3, -0.2, -0.1, 0.0 \\ 
\hline
\multicolumn{2}{c}{Dust emission \citep{Dale2014}} \\
\hline
$\alpha$ & 0.25, 0.5, 1.0, 1.5, 2.0, 2.5, 3.0, 3.5, 4.0 \\
\enddata
\end{deluxetable}

\subsection{Spatially-resolved photometry and SED fitting} \label{sec:resolvedmaps}

To measure the spatially-resolved photometry, we use the PSF-matched imaging and reproject them to a pixel scale of 0\farcs1, to match the \ha\ emission line map. 
To place stronger constraints on the resolved properties, we applied an adaptive binning algorithm\footnote{\url{https://github.com/pierrethx/MVT-binning}} based on the Weighted Voronoi Tessellation (MVT) method described in \citep{Cappellari2003, Diehl2006}. 
We chose 2 pixels to be the minimum size of the Voronoi bin, which is approximately the size of the FWHM of the PSF of F160W (FWHM = $0\farcs18$). 
We run \cigale\ using the 21-band fluxes in each Voronoi bin, and the same modules and parameters as described in Section \ref{sec:sed}. 

As a check on the accuracy of the results from the spatially resolved SED fitting, we compared the sum of the stellar mass and SFR$_\mathrm{100~Myr}$ values from all of the spatially resolved bins within a galaxy to the values derived from SED fitting to the integrated photometry from the whole galaxy (where the latter includes longer-wavelength data, if available, see above).
Figure~\ref{fig:integrated_vs_resolved} shows this comparison. 
The stellar mass and $\mathrm{SFR_{100Myr}}$ summed from the spatially-resolved bins and from the integrated SED fitting agree well with each other. Formally, we measure small median differences of 0.06 dex and -0.03 dex in stellar mass and SFR$_\mathrm{100Myr}$, respectively.   

\subsection{\ha\ Equivalent Width Maps} \label{sec:ew}

We use the H$\alpha$ maps from \grizli\ reconstructed from the NIRISS WFSS data for each galaxy. From these maps, we calculated the rest-frame EW(\ha) maps as follows, 
\begin{equation}
    EW(\ha) = \frac{f_\mathrm{H\alpha}}{f_\mathrm{continuum}}\ \frac{1}{(1+z)},
\end{equation}
where we take the continuum flux $f_\mathrm{continuum}$ to be,
\begin{equation}\label{eqn:f_cont}
    f_\mathrm{continuum} = \frac{f_\mathrm{direct-image} \times \Delta\lambda - f_\mathrm{H\alpha}}{ \Delta\lambda}.
\end{equation}
In these equations, $f_\mathrm{H\alpha}$ is the flux from the \ha\ emission-line map, 
%
$f_\mathrm{direct-image}$ is flux from the NIRISS direct imaging in the bandpass that contains the \ha\ line, 
$\Delta\lambda$ is the FWHM of the direct-image bandpass, and $(1+z)$ is a factor to correct to the rest-frame. 
The uncertainty on EW(\ha) is obtained following the propagation of uncertainty using the associated error maps. 
To obtain secure EW(\ha) measurements, we include only those pixels where the \ha\ flux has significance $> 3\sigma$, and the underlying continuum flux from the direct image has significance $>5\sigma$. 

Due to the low spectral resolution of the NIRISS grism, the \ha\ and \nii\ lines are blended. 
Therefore, in this work, EW(\ha) includes the sum of the \ha\ and \nii\ lines (For simplicity, we use EW(\ha) in the text, but use ``EW(\ha+\nii)'' in the figure for clarity). 
For the stellar mass and redshift of the galaxies in our sample, we expect \nii/(\ha+\nii) to be $\sim 0.2$ \citep{Faisst2018}. 
Because we are looking at radial gradients in the EW(\ha), this has no impact on our results so long as there are no strong changes in \ha+\nii\ throughout the galaxies {(see further discussion in Section \ref{sec:meta})}. 
{Another effect on the} EW(\ha) measurements is impacted by the presence of AGN, which could have higher \ha+\nii\ values \citep{Baldwin1981}. 
This was the motivation to remove the inner $3\times 3$ pixels of galaxies that show AGN activity (see above).

We also compare the EW(\ha) from the spatially resolved measurements against 
{the EW(\ha) measured from the best-fitted 1D integrated spectrum. The 1D spectrum is extracted from the stacked 2D spectra using an optimally-weighted method following \citet{Horne1986}. The integrated EW(\ha) is calculated using }

\begin{equation}
    EW(\ha)_{\mathrm{integrated}} = \int \frac{F_{\mathrm{H}\alpha}}{F_{\mathrm{continuum}}} \,d\lambda_{\mathrm{rest}}
\end{equation}
{where $F_{\mathrm{H}\alpha}$ is the best-fitted \ha\ emission line profile, and $F_{\mathrm{continuum}}$ is obtained by fitting a linear function to the best-fitted spectra in two wavelength ranges 6510\AA\ $< \lambda_{rest} <$ 6550\AA\ and 6620\AA\ $< \lambda_{rest} <$ 6660\AA. These two wavelength windows are chosen to be close to the \ha+\nii\ regions but are not affected by the Balmer absorption. The uncertainty is obtained from a Monte-Carlo where we perturb each bin using the full covariance matrix of the best-fitted spectra. }

{The average EW(\ha) from the emission line flux maps is calculated using }
\begin{equation}
   \langle EW(\ha) \rangle = \frac{1}{(1+z)} \times \frac{ \sum_i f_{i,\mathrm{H}\alpha}}{\sum_i f_{i,\mathrm{continuum}}} 
\end{equation}
where $f_{i,\mathrm{H}\alpha}$ are the \ha\ fluxes in each pixel and $f_{i,\mathrm{continuum}}$ are the continuum fluxes in each pixel from the direct imaging.  The sum is over all $i$ pixels that have EW(\ha) measurements. 
{The uncertainty on $\langle EW(\ha) \rangle$ is obtained using the associated error maps. We note that the uncertainties on the integrated EW(\ha) and for $\langle EW(\ha) \rangle$ from the maps do not include uncertainties for the background subtraction and contamination uncertainties.  The former should be negligible assuming the relative uncertainties are small on the size of our galaxies.  Our inspection of the images suggests the latter is also negligible, but we can test this with the second epoch of NGDEEP NIRISS data.  }

\begin{figure*}
    \includegraphics[width=\textwidth]{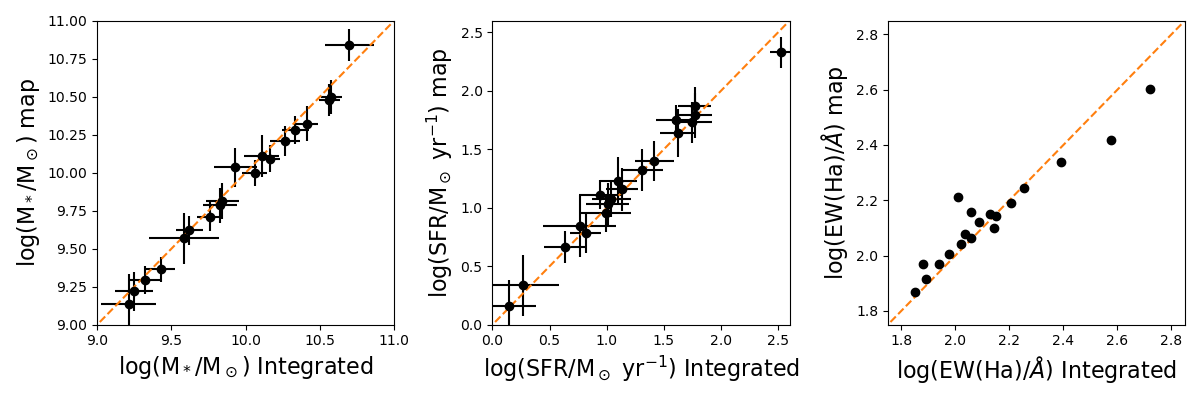}
    \caption{\textit{Left and Middle:} Comparison between the stellar mass and star-formation rate derived by summing the values from the bins in the spatially-resolved ``map(s)'' with the same aperture as used for integrated photometry and the ``integrated'' values derived from integrated photometry. \textit{Right:} Comparison between the \ha\ equivalent width average from the EW(\ha) ``map(s)'' and the ``integrated'' EW(\ha) measured by \grizli\ from the total 1D spectra. Note that statistical errors on the EW(\ha) values from the maps and from \grizli\ are smaller than the data points. }
    \label{fig:integrated_vs_resolved}
\end{figure*}

The results are shown in {the right panel of} Figure~\ref{fig:integrated_vs_resolved}.  
{We see the majority of data follows the 1--to--1 line, with a few outliers. }
The median difference is {0.02 dex with the 16th/84th percentiles of the distribution of differences are $-$0.05/0.04 dex}. 

{We consider here some possible reasons that could cause the differences between the integrated EW(\ha) and that calculated from the resolved maps. 
The integrated EW(\ha) is calculated from the 1D spectra, and the extraction from the stacked 2D spectra uses the optimally weighted method. We calculate the $\langle EW(\ha) \rangle$ from the resolved maps using an unweighted average over the pixels that have EW(\ha) detections. Secondly, the 1D spectra are stacked from spectra taken in two diagonal PAs, (i.e., GR150R and GR150C), there can be issues where resolved structures are poorly modeled by Grizli, which models all features as a single-Gaussian model. This could lead to an underestimate of the fluxes in some galaxies. Thirdly, for the $\langle EW(\ha) \rangle$ from the resolved maps, }
it is also possible that for more dusty and redder galaxies where the continuum is rising toward redder wavelengths, our assumption of a  flat continuum assumption could potentially bias the EW(\ha) and depend on the observed-frame wavelength of \ha\ relative to the bandpass. 
{In addition, there are differences in how the stellar absorption is accounted for in the EW measurements. \grizli\ includes an estimate of the stellar continuum when it generates the emission-line flux images 
Here, we then use the direct image to model the underlying continuum (Equation~\ref{eqn:f_cont}).  Because the direct imaging includes the effects of the stellar continuum, we are 
double-subtracting the continuum from the emission line fluxes. This could underestimate the underlying continuum, and overestimate the resolved the $\langle EW(\ha) \rangle$.  
We tested the effects of stellar absorption in the continuum when measuring the EW by comparing the EW(\ha) measured using the continuum correction from \grizli\ compared to using a linear fit to the continuum (see above).  We measure the differences in the range of 0.06 - 0.08 dex in the integrated EW(\ha) between using a linear function as an underlying continuum and using a continuum with absorption. This is an upper limit because the continuum from the direct image is averaged over a longer wavelength, instead of averaging over the \ha\ region as in the integrated EW(\ha). Therefore, we take this as an upper limit on the systematic uncertainty in the EW arising from how we model the continuum. In addition, the effect of other emission lines falling in the same filter is negligible. After removing other emission lines, the $\langle EW(\ha) \rangle$ is lower by a range of 0.01 - 0.03 dex. }
{We note that} these differences could affect our results when we compare them to integrated studies (Section~\ref{sec:ew}), but this does not change our results on the EW(\ha) gradients. 


\section{Results} \label{sec:results}

\subsection{Spatially Resolved Equivalent Widths, specific SFRs, and Ages} \label{sec:ew}

\begin{figure*}
    \centering
    \includegraphics[width=\textwidth]{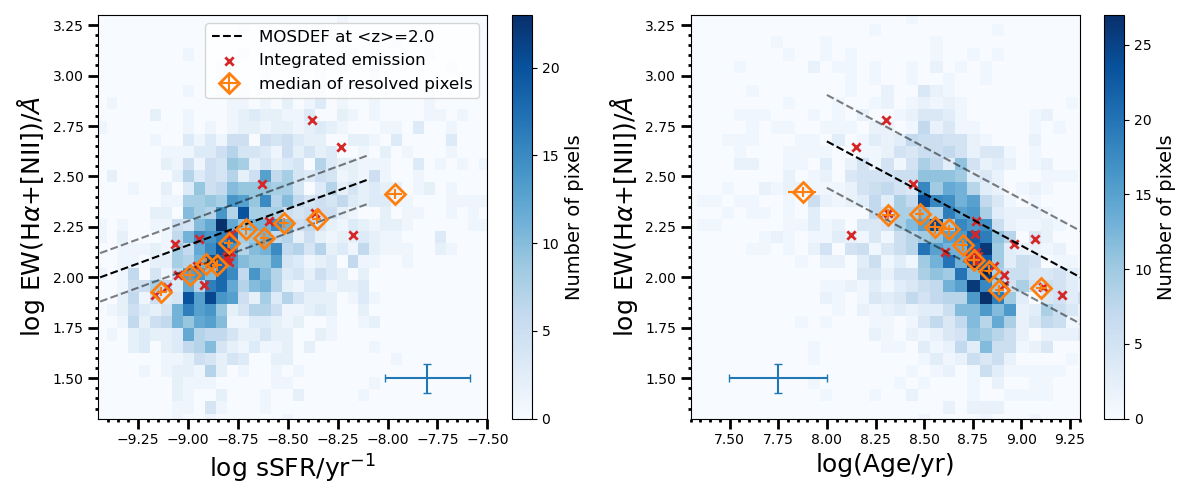}
    \caption{Relation between spatially resolved sSFR and EW(\ha) (\textit{left}), and spatially resolved age and EW(\ha) (\textit{right}).  The 2D histograms include all pixels in the spatially resolved images for galaxies in our sample. In each panel, the orange dots show the medians of the values and errors derived from the individual pixels. The {red} crosses show values derived from the integrated photometry. The median error of individual, spatially resolved pixels is indicated by the blue error bar in each panel. For comparison, the dashed lines show the relations derived from MOSDEF galaxies at $z=1.4-3.8$ \citep{Reddy2018} (median and 1$\sigma$ scatter). The median values of our spatially resolved measurements broadly agree with the results from the integrated emission of our galaxies. However, these are offset slightly to lower EW(\ha) compared to the relation from MOSDEF. }
    \label{fig:bins}
\end{figure*}

The analysis of the integrated emission of galaxies has shown strong correlations between the equivalent widths of various nebular emission lines, the specific SFR (sSFR), and the age of the stellar population from $z\sim 0$ to 6 \citep{Schaerer2013, Reddy2018}.  Here, we use the spatially resolved information in the \jwst\ and \hst\ data to study these correlations within individual galaxies. 

Figure \ref{fig:bins} shows the 2D histograms of all pixels from the spatially resolved maps that include sSFR and age versus EW(\ha). We also show the average (median) values in bins along the abscissa. The median and associated median error are calculated with a bootstrap method following \citet{Shen2023}. Each median and its error are calculated using approximately a similar number of data points ($\sim$200 individual pixels) per median. 

Focusing on the spatially resolved results, the EW(\ha) increases with increasing sSFR and decreases with increasing age. We adopt the non-parametric Spearman's $\rho$ test to assess these correlations\footnote{The Spearman's $\rho$ test returns a correlation coefficient $\rho$ and associated $p$-value. The $p$-value quantifies the significance of the correlation by giving the probability that the data are uncorrelated. We reject the null hypothesis for $p$-value $\leq$ 0.05.}. The returned $p$-values are much smaller than 0.01, therefore, we \textit{reject} the null hypothesis that the EW(\ha) and SSFR and age are \textit{uncorrelated}. Furthermore, we note that the EW(\ha) and sSFR (and EW(\ha) and age) relations are independent measurements because the EW(\ha) is obtained from the NIRISS grism data independent of the SED fitting results. In contrast, the sSFR and age are derived from SED fitting to the broad-band and medium-band photometry. 

We compare our trends derived from the spatially resolved maps to those derived from integrated measurements.  For the integrated properties of galaxies, we use the sSFR (and age) versus EW(\ha+\nii) relations for star-forming galaxies at $z=1.4-3.8$  from the MOSFIRE Deep Evolution Field (MOSDEF) survey \citep{Reddy2018}. 
Our median sSFR versus EW(\ha) and mass-weighted age versus EW(\ha) trends are offset from those of MOSDEF  by a median of 0.11 dex and 0.18 dex, respectively. This is comparable to the 1$\sigma$ scatter reported in MOSDEF by \citet{Reddy2018}. {This offset could be due to the different star formation histories adopted in the SED fitting that MOSDEF assumed a constant star formation history. }

Furthermore, we see a clear curvature in our resolved median sSFR--EW(\ha) and age--EW(\ha) values.  Specifically, the slope of the relations is ``flatter'' for high sSFR and young ages. This is also apparent in the integrated relations. 
In addition, the sSFR versus EW(\ha) and age versus EW(\ha) relations show a large scatter. Quantitatively, the scatter is {0.30}~dex and {0.29}~dex in the sSFR versus EW(\ha) relation and age versus EW(\ha) relation, respectively. This scatter is larger than that reported from the MOSDEF derived from the integrated properties of galaxies, where \citet{Reddy2018} found 0.12 dex and 0.23 dex, respectively. 
We do not consider these to be significant given the small number of objects in our sample, and the different methods used by MOSDEF and here.  
Instead, we suspect that the scatter and the shape of our resolved sSFR and age versus EW(\ha) reflect complex evolutionary trends inside galaxies and galaxy-to-galaxy variation, while the trends of integrated measures only reflect the latter. 
We will discuss this further in Section~\ref{sec:discussion}. 

Regardless of the systematic offsets and scatter, these comparisons indicate a consistent relation between the sSFR and age versus EW(\ha) relations for both the integrated emission from galaxies and the spatially-resolved properties.  
This means that these trends are driven by star formation and stellar populations within galaxies on spatial scales on the order of kiloparsecs.

\subsection{Radial Gradients in EW(\ha), sSFR and Age} \label{sec:gradients}

\begin{figure*}
    \centering
    \includegraphics[width=0.95\textwidth]{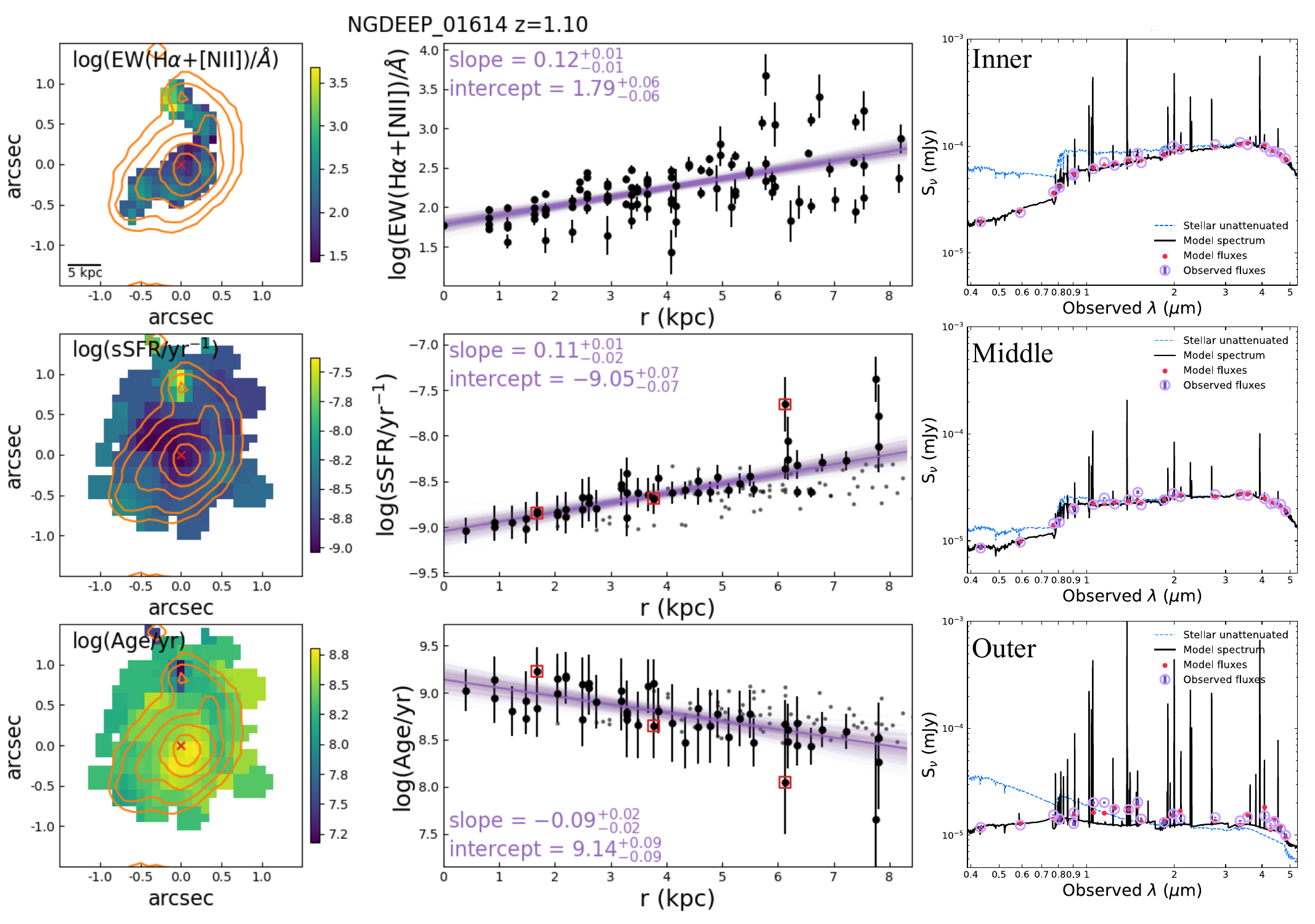}
    \caption{Example of the EW(\ha), sSFR, and age spatial maps and their radial gradients for galaxies in our sample. {The sSFR and age maps use the} Voronoi tessellation (see Sect. \ref{sec:resolvedmaps}). In the left panels, the red crosses mark the luminosity-weighted galaxy centroid derived from the F160W image. Orange contours are 5, 10, 25, 50, 100 $\sigma$ in the F160W image. The {example here shows a galaxy with no AGN indication (NGDEEP\_ 01614)}. The {middle} panels show the radial gradient of the EW(\ha), sSFR, and age. 
    %
    %
    The purple solid lines show best-fit linear relations, along with 400 random draws from the posterior.
    Grey dots in the sSFR and age panels show Voronoi bins that lack EW(\ha) measurements, which are excluded in the linear fitting. {The right panels show SED fits at three different radial bins (``inner'', ``middle'', and ``outer'' regions, top-to-bottom), indicated by the red boxes in the middle panels. Each panel shows the observed photometric fluxes with errors (purple), the CIGALE-derived best-fitted model (black curve), and model photometry (red dots). The best-fitted CIGALE model is the sum of the contributions from a dust-attenuated stellar emission and nebular emission (the unattenuated stellar model is shown in blue). }{A galaxy with an AGN is shown in Figure~\ref{fig:radialprofiles2}.  The maps and radial gradients of the remainder of our galaxies are shown in Appendix \ref{app:maps}. }}
    \label{fig:radialprofiles}
\end{figure*}

\begin{figure*}
    \centering
    \includegraphics[width=0.95\textwidth]{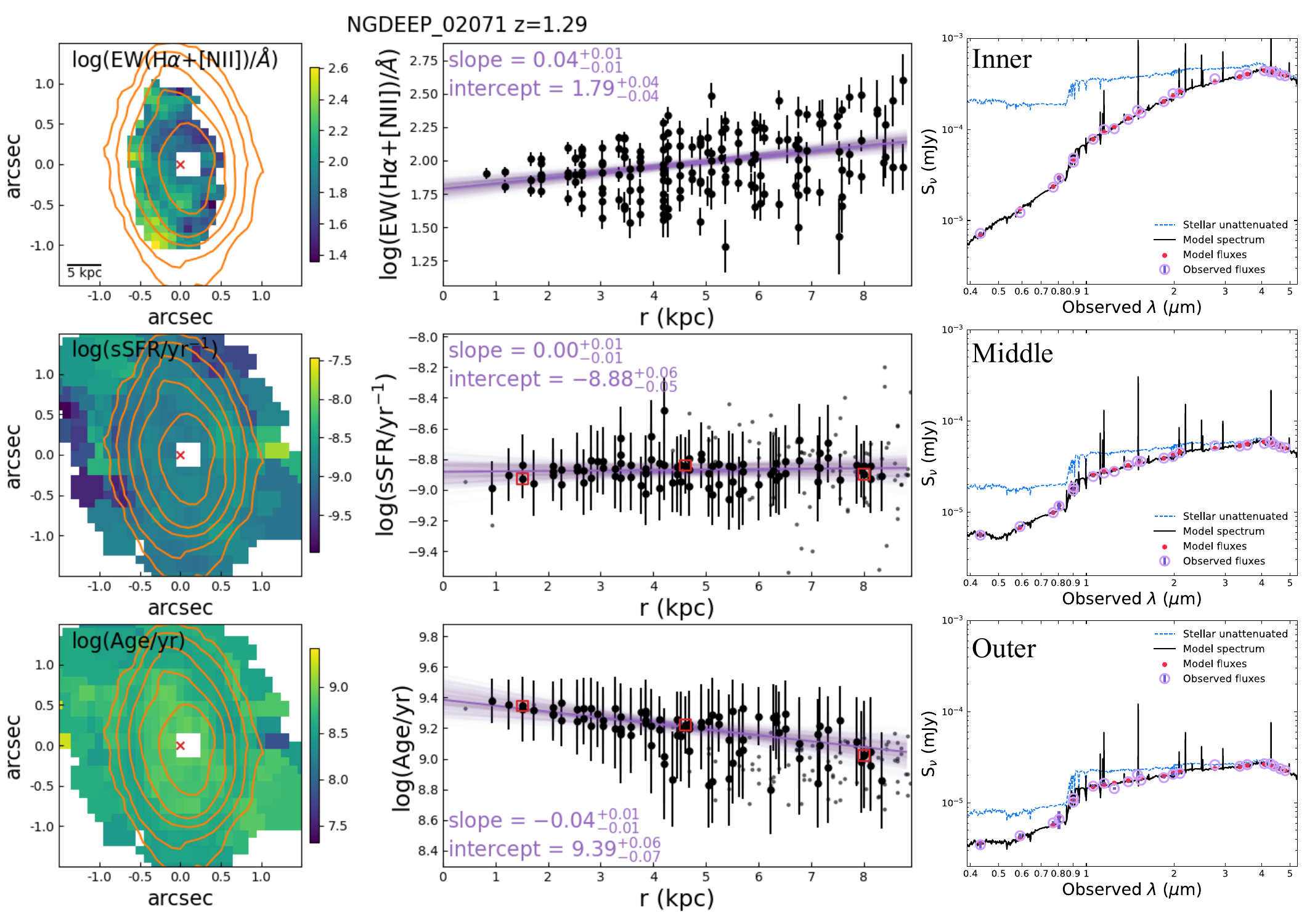}
    \caption{{Same as Figure~\ref{fig:radialprofiles} but for a galaxy flagged with an AGN (NGDEEP\_02071). For all galaxies with AGN, we mask the central $3x3$ pixels to remove possible contamination (see Section~\ref{sec:sample}, as seen in this galaxy).  The maps and radial gradients of the remainder of our galaxies are shown in Appendix \ref{app:maps}. } }\label{fig:radialprofiles2}
\end{figure*}

\begin{figure*}
    \centering
    \includegraphics[width=\textwidth]{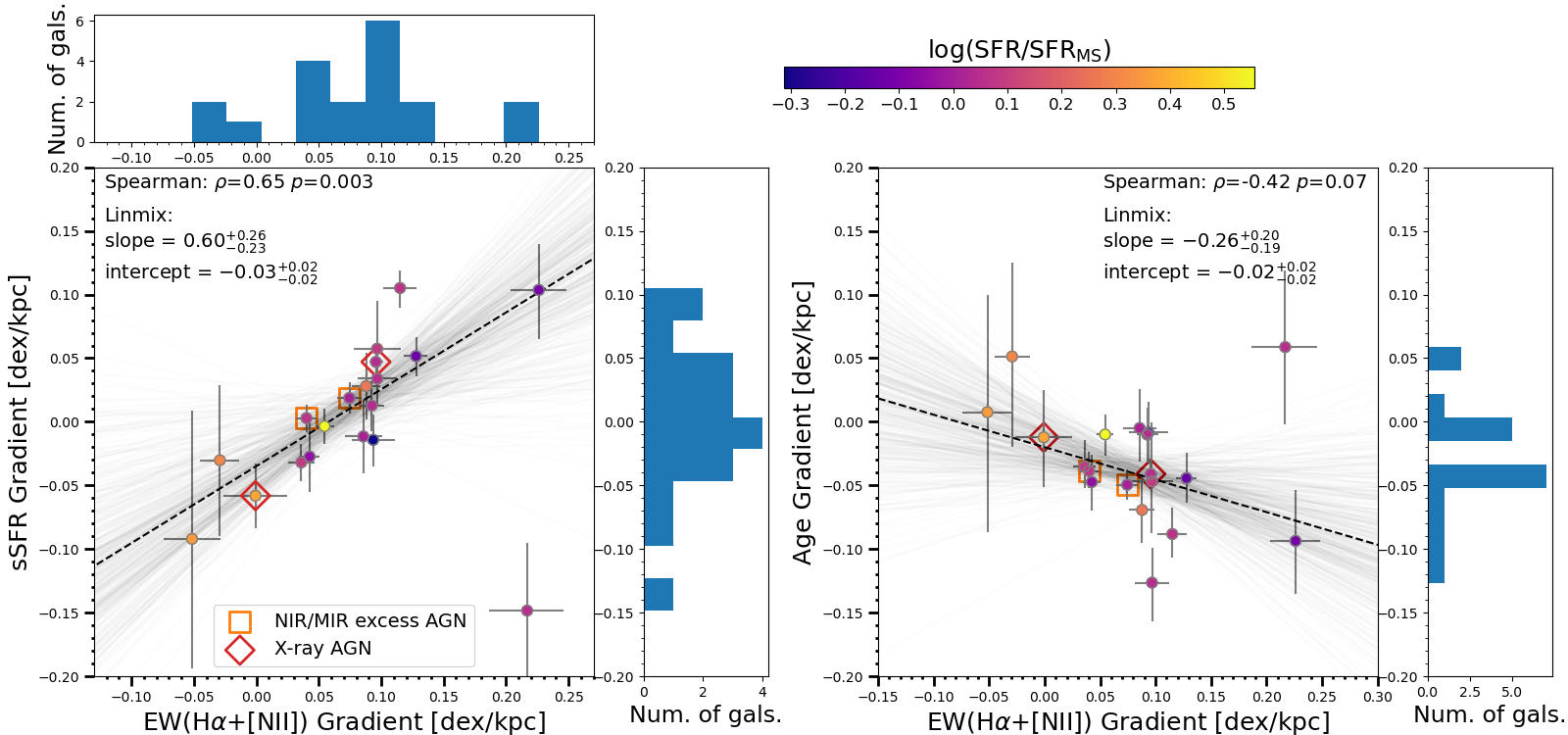}
    \caption{The radial gradients of EW(\ha) versus sSFR (\textit{left}), and the radial gradients of EW(\ha) versus stellar mass-weighted age (\textit{right}).  Here the gradients are the slope measured for the radial gradients for each galaxy (see Figure~\ref{fig:radialprofiles}). In each plot, the histograms show the distribution of EW(\ha), sSFR, and age gradients for the individual galaxies. The markers in the scatter plots are color-coded by the distance of the galaxies from the SFMS (positive/negative values are above/below the SFMS). Galaxies identified with AGN are marked by open red diamonds and orange squares. The best-fit linear relations are shown as the purple dash lines along with 400 random draws from the posterior (faint grey lines). }
    \label{fig:compare_gradient}
\end{figure*}

The radial profiles of EW(\ha), sSFR, and age can test the stellar build-up and quenching processes.  For example, different radial trends are expected if these processes proceed in an ``inside-out'' or ``outside-in'' fashion (e.g., \citealp{Ellison2018}). 
To quantify the radial gradient, we fit for each galaxy a linear relation to the logarithm of EW(\ha), sSFR, and age as a function of the proper distance to the luminosity-weighted center (the ``galactocentric radius'').  We use a Gaussian Mixture Model method (\linmix, \citealp{Kelly2007}) to fit the relation, which allows for uncertainties on the dependent variables (EW(\ha), sSFR, and age). The best-fit parameters are the median of the fitted parameters (the slope and intercept) from 400 random draws from the posterior, and the associated errors are the 16th and 84th percentiles of each parameter. 
We reiterate that for galaxies with AGNs, their central pixels ($0\farcs3\times0\farcs3$) are not included in the fitting. However, if these pixels are included, we do not find a significant change in the measured gradients.   
In Figure~\ref{fig:radialprofiles}, we show two examples of the EW(\ha), sSFR, and age maps, and as a function of galactocentric radii with their best-fitted lines. 
The radial gradients of EW(\ha), sSFR, and age for our galaxies are listed in Table~\ref{tab:props}. 
{In the Appendix we show the EW(\ha), sSFR, and age maps, and their radial profiles for the rest of the galaxies in our sample.
}

\begin{deluxetable*}{cccccccccc}
\rotate
\tablecolumns{10}
\tablecaption{Galaxy properties \label{tab:props} }
\tablehead{ \colhead{ID} & \colhead{R.A.} & \colhead{Decl.} & \colhead{$z_{\rm spec}$} & \colhead{$\mathrm{log(M_*/M_\odot)}$} & \colhead{$\mathrm{log(SFR/M_\odot~yr^{-1})}$} & \colhead{EW(\ha)}& \colhead{EW(\ha) gradient} & \colhead{sSFR gradient} & \colhead{age gradient}\\ [-4pt]
\colhead{} & \colhead{(J2000)} & \colhead{(J2000)} & \colhead{} & \colhead{} & \colhead{} & \colhead{\AA} & \colhead{dex~kpc$^{-1}$} & \colhead{dex~kpc$^{-1}$} & \colhead{dex~kpc$^{-1}$}\\ [-4pt]
\colhead{(1)} & \colhead{(2)} & \colhead{(3)} & \colhead{(4)} & \colhead{(5)} & \colhead{(6)} & \colhead{(7)} & \colhead{(8)} & \colhead{(9)} & \colhead{(10)} }
\startdata
NGDEEP\_00862 & 53.16804 & -27.78974 & 0.619 & 9.25$\pm$0.13   & 0.14$\pm$0.23 & 75.6$\pm$2.2 & $0.226_{-0.022}^{+0.023}$ & $0.104_{-0.036}^{+0.038}$ & $-0.094_{-0.040}^{+0.042}$ \\ 
NGDEEP\_00922 & 53.14721 & -27.78855 & 0.667 & 9.21$\pm$0.19   & 0.27$\pm$0.32 &  133.9$\pm$1.5 & $0.217_{-0.029}^{+0.030}$ & $-0.148_{-0.053}^{+0.058}$ & $0.059_{-0.061}^{+0.061}$ \\
NGDEEP\_02086 & 53.15526 & -27.76961 & 0.736 & 9.93$\pm$0.15   & 0.76$\pm$0.32  & 71.2$\pm$3.4 & $0.043_{-0.008}^{+0.008}$ & $-0.027_{-0.030}^{+0.028}$ & $-0.047_{-0.022}^{+0.023}$ \\
NGDEEP\_01614 & 53.17299 & -27.77794 & 1.096 & 9.43$\pm$0.10  & 0.64$\pm$0.18  & 114.1$\pm$2.4 & $0.115_{-0.013}^{+0.014}$ & $0.105_{-0.014}^{+0.015}$ & $-0.088_{-0.020}^{+0.018}$ \\
NGDEEP\_00500 & 53.15589 & -27.79497 & 1.096 & 9.59$\pm$0.24   & 0.99$\pm$0.22  & 161.2$\pm$2.9 & $0.088_{-0.011}^{+0.012}$ & $0.028_{-0.026}^{+0.027}$ & $-0.069_{-0.028}^{+0.026}$ \\
NGDEEP\_01268 & 53.17824 & -27.78315 & 1.119 & 9.62$\pm$0.09  & 0.82$\pm$0.14  & 104.6$\pm$1.5 & $0.097_{-0.016}^{+0.015}$ & $0.034_{-0.022}^{+0.020}$ & $-0.126_{-0.028}^{+0.030}$ \\
NGDEEP\_01837 & 53.17607 & -27.77379 & 1.288 & 10.26$\pm$0.10 & 1.31$\pm$0.18  & 94.8$\pm$2.8 & $0.074_{-0.010}^{+0.010}$ & $0.019_{-0.012}^{+0.012}$ & $-0.049_{-0.012}^{+0.012}$ \\
NGDEEP\_02071 & 53.16633 & -27.76867 & 1.295 & 10.33$\pm$0.09 & 1.42$\pm$0.17  & 77.9$\pm$1.6 & $0.040_{-0.008}^{+0.008}$ & $0.003_{-0.011}^{+0.011}$ & $-0.039_{-0.015}^{+0.013}$ \\
NGDEEP\_01974 & 53.14619 & -27.77111 & 1.316 & 10.06$\pm$0.09 & 1.01$\pm$0.19  & 87.3$\pm$1.6 & $0.128_{-0.009}^{+0.010}$ & $0.052_{-0.015}^{+0.016}$ & $-0.044_{-0.020}^{+0.019}$ \\
NGDEEP\_00891 & 53.14510 & -27.78950 & 1.320 & 9.83$\pm$0.12    & 1.04$\pm$0.17  & 141.3$\pm$2.0 & $0.036_{-0.010}^{+0.010}$ & $-0.032_{-0.014}^{+0.015}$ & $-0.035_{-0.021}^{+0.017}$ \\
NGDEEP\_00997 & 53.16170 & -27.78750 & 1.836 & 10.58$\pm$0.07 & 1.77$\pm$0.15   & 108.6$\pm$3.0 & $0.093_{-0.010}^{+0.010}$ & $0.013_{-0.020}^{+0.020}$ & $-0.010_{-0.021}^{+0.022}$ \\
NGDEEP\_01520 & 53.15568 & -27.77934 & 1.846 & 10.56$\pm$0.07 & 1.77$\pm$0.15  & 102.1$\pm$1.2 & $0.096_{-0.006}^{+0.005}$ & $0.047_{-0.011}^{+0.011}$ & $-0.041_{-0.011}^{+0.010}$ \\
NGDEEP\_02088 & 53.15231 & -27.77016 & 1.848 & 9.76$\pm$0.09 & 1.13$\pm$0.14    & 245.8$\pm$3.4 & $0.097_{-0.019}^{+0.019}$ & $0.057_{-0.038}^{+0.036}$ & $-0.047_{-0.047}^{+0.041}$ \\
NGDEEP\_01545 & 53.14924 & -27.77887 & 1.850 & 10.17$\pm$0.07 & 1.10$\pm$0.16   & 122.1$\pm$4.8 & $0.094_{-0.017}^{+0.017}$ & $-0.014_{-0.020}^{+0.021}$ & $-0.009_{-0.024}^{+0.023}$ \\
NGDEEP\_01831 & 53.14419 & -27.77363 & 1.900 & 9.33$\pm$0.10 & 0.94$\pm$0.18     & 523.5$\pm$9.3 & $-0.052_{-0.022}^{+0.023}$ & $-0.092_{-0.101}^{+0.101}$ & $0.008_{-0.092}^{+0.094}$ \\
NGDEEP\_00923 & 53.14932 & -27.78859 & 1.906 & 10.11$\pm$0.12 & 1.74$\pm$0.17   & 178.8$\pm$1.5 & $-0.001_{-0.026}^{+0.025}$ & $-0.058_{-0.025}^{+0.025}$ & $-0.012_{-0.037}^{+0.039}$ \\
NGDEEP\_01346 & 53.17433 & -27.78259 & 1.998 & 10.41$\pm$0.08 & 1.62$\pm$0.16  & 114.7$\pm$2.3 & $0.086_{-0.015}^{+0.015}$ & $-0.012_{-0.029}^{+0.029}$ & $-0.005_{-0.030}^{+0.027}$ \\
NGDEEP\_00272 & 53.16692 & -27.79880 & 2.000 & 10.70$\pm$0.17 & 2.52$\pm$0.10   & 139.3$\pm$1.6 & $0.055_{-0.007}^{+0.007}$ & $-0.004_{-0.014}^{+0.014}$ & $-0.010_{-0.016}^{+0.017}$ \\
NGDEEP\_01948 & 53.15448 & -27.77151 & 2.226 & 9.84$\pm$0.11 & 1.61$\pm$0.18    &  375.9$\pm$3.3 & $-0.029_{-0.016}^{+0.015}$ & $-0.030_{-0.059}^{+0.059}$ & $0.051_{-0.074}^{+0.071}$ \\
\enddata
\tablecomments{ (1-4) The ID, coordinate, and redshift of our sample. (5) and (6) The stellar mass and SFR and associated uncertainties are from integrated SED fitting (see Section~\ref{sec:sed}). (7) The integrated EW(\ha) derived from the 1D spectrum from \grizli. (8-10) The radial gradients of EW(\ha), sSFR, stellar mass-weighted age, and associated uncertainties measured from their resolved maps, as described in Section~\ref{sec:gradients}.  }
\end{deluxetable*}

Figure~\ref{fig:compare_gradient} compares the radial EW(\ha) gradients with the sSFR and age gradients. The majority of galaxies in our sample (16/19) have positive EW(\ha) gradients (i.e., the EW(\ha) increases with increasing galactocentric radius).  The median of EW(\ha) gradients is {$0.09^{+0.03}_{-0.06}$} dex/kpc, where the uncertainties are the 16th/84th percentiles. Interestingly, the two galaxies (NGDEEP\_01948 and NGDEEP\_01831) with the most negative EW(\ha) gradients show distinct morphological features indicative of mergers or large star-forming ``clumps'', which impacts the gradients (see also Fig. \ref{fig:sample}). We discuss this further below (see Section \ref{sec:discussion}). 

For the sSFR and age gradients, there are {10 and 16} galaxies, respectively, that have positive sSFR gradients (sSFR increases with galactocentric radius) and negative age gradients (age decreases with radius). 
The median sSFR and age gradients are {$0.003^{+0.049}_{-0.038}$} dex/kpc and {$-0.04^{+0.04}_{-0.03}$} dex/kpc. 

There is a correlation between the strength of the EW(\ha) gradient and that of the sSFR gradient in the galaxies in our sample. The Spearman's $\rho$ test applied to the data in Figure~\ref{fig:compare_gradient} returns a $p$-value of {0.003} which we interpret as evidence ($>2.5\sigma$) that a correlation exists.  
While there is an \textit{anti-}correlation between the strength of the EW(\ha) gradient and that of the age gradient (right panel of Figure~\ref{fig:compare_gradient}, the Spearman's $\rho$ test ($p$-value={0.07}) result is less conclusive. This could be a result of different factors, which we discuss below in Section \ref{sec:discussion}, but at least some of the \textit{weakness} in the correlations is because of one galaxy that ``bucks the trend'', which we discuss in the next paragraph.

\begin{figure}
    \centering
    \includegraphics[width=\columnwidth]{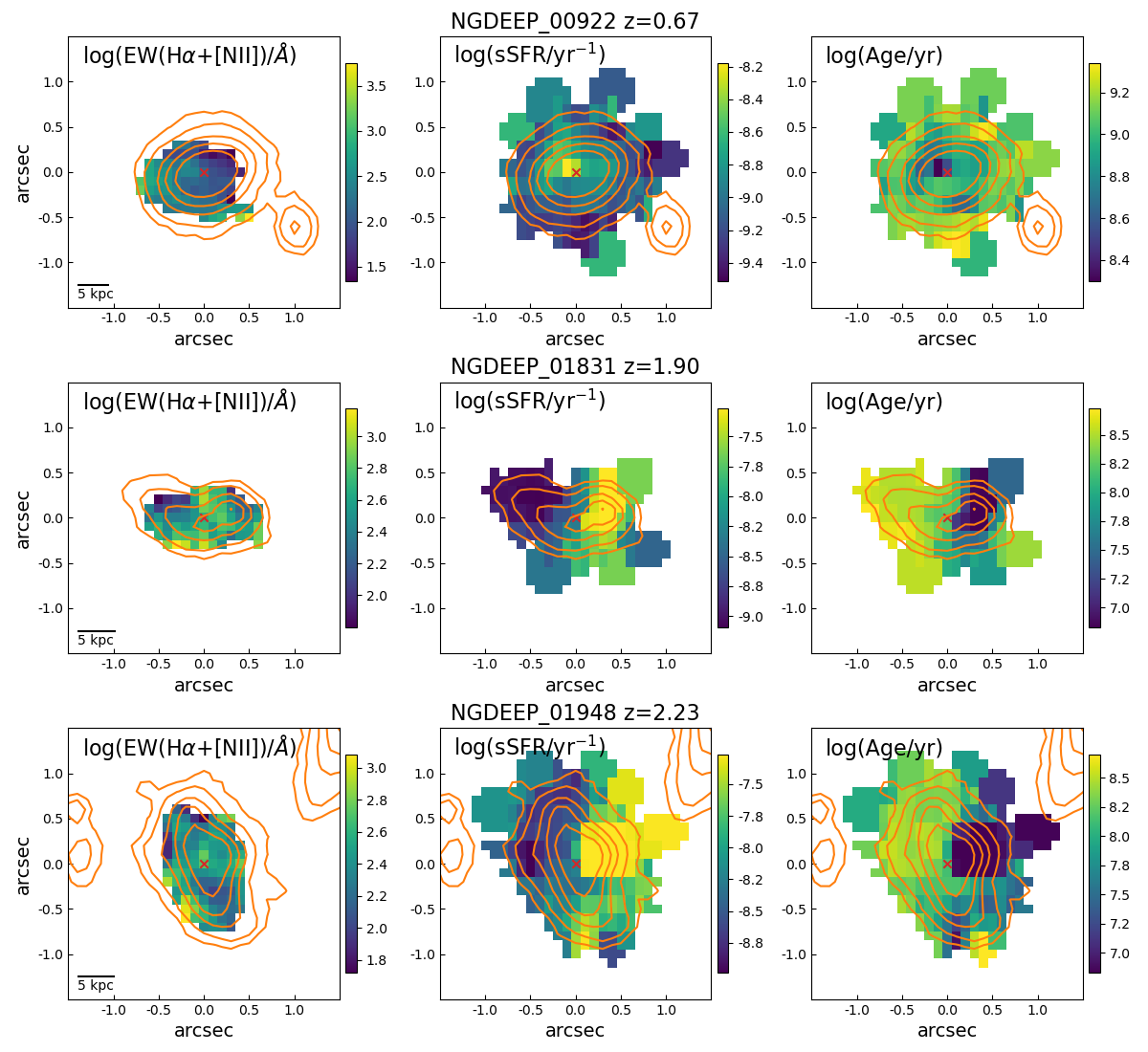}
    \caption{The EW(\ha), sSFR, and age spatial maps for galaxies with atypical gradients.  The top row shows NGDEEP\_00922, which shows a large offset from the relationship in both EW(\ha) -- sSFR and -- age gradients (Section~\ref{sec:gradients}).  The middle and bottom rows show the two galaxies with negative EW(\ha) gradients (NGDEEP\_01831 and NGDEEP\_01948; Section~\ref{sec:negative_EW}) }
    \label{fig:maps_for3}
\end{figure}

There is one galaxy, NGDEEP\_00922, that shows a large offset from the relationship in both EW(\ha) versus sSFR and EW(\ha) versus age gradients in Figure~\ref{fig:compare_gradient}. This galaxy has a high EW(\ha) gradient of 0.22~dex kpc$^{-1}$ but very negative sSFR gradient of $-0.14$~dex kpc$^{-1}$. 
The reason for this is that this galaxy's sSFR and age maps show a region of high sSFR and low age region near the center (the top row of Figure~\ref{fig:maps_for3}), while this feature does not appear with an excess in the  EW(\ha) map.   This galaxy may have either very recently quenched the star formation in the core (leading to a drop in EW(\ha)) that has yet to appear in the broad-band SED.  It is also possible that the degeneracy between the age and dust attenuation in the broad-band SED fits have misrepresented the physical conditions in this bin (leading to an erroneously high sSFR). This could be tested by using spatially resolved mid-IR imaging (e.g., \jwst/MIRI), which will be available in the future.  Regardless, if we remove this one object from our fits, the correlation between the EW(\ha) gradient and sSFR gradient becomes even stronger, where the Spearman $\rho$ test returns a $p$-value of {$10^{-7}$}.  Similarly, the \textit{anti-}correlation between the EW(\ha) gradient and age gradient becomes stronger, where this test returns a $p$-value of {0.004}. 

Previous studies have argued that the shape of the sSFR radial profile correlates with a galaxy's distance from the center of the SFMS.  This has been reported at low-redshift \citep{Ellison2018}, at $z\sim1$ \citep{Nelson2021} and at $z\sim2$ \citep{Tacchella2018}. 
We therefore consider if our galaxies follow this trend.  We color-coded galaxies based on the distance from the SFMS, using the ratio between the measured SFR and SFR on the SFMS relation at their stellar mass and redshift ($\mathrm{ \log(SFR/SFR_{MS})}$).  This is shown in Figure~\ref{fig:compare_gradient} where positive (negative) values indicate galaxies with SFRs above (below) the SFMS.  
Qualitatively, the gradients of EW(\ha) and sSFR tend to be more positive for galaxies below the main sequence, and more negative for galaxies above the main sequence. 
We do not observe such a trend in the age gradients. 
However, we cannot make any quantitative conclusions based on the small sample of 19 objects that span a wide range of stellar mass, SFR, and redshift. We plan to return to this question once larger samples are available. 

\section{Discussion} \label{sec:discussion}

{Previous studies have seen that the integrated (total) specific sSFRs and gas fractions are correlated with galaxy structure \citep[e.g.,][]{Papovich2015, Whitaker2015, Freundlich2019}.  This has been observed also in studies of spatially resolved galaxies  \citep[e.g,][]{Wuyts2011, Wuyts2013}.  The implication is that galaxy quenching and star-formation activity are linked.  Here we see that this process is tied to the sSFR and EW(\ha) values within a galaxy. }

The majority of our sample (16 out of 19 {galaxies}) show positive EW(\ha) gradients {where ``positive'' means the EW increases with radius. The} median EW(\ha) gradient {is 0.09} dex~kpc$^{-1}$. This implies there is active build-up {of the stellar content} in the outer regions of galaxies, compared to the inner-most regions. 
This result supports an ``inside-out'' scenario of galaxy disk formation, where the central component is built first (and quenches first) and the disk structure grows radially, forming stars at preferentially larger radii  (e.g., \citealp{Cole2000, vandenBosch2002, Papovich2005, Aumer2013, Somerville2015}). 
Central suppressed {star formation has been seen in the interpretation of EW(\ha) gradients} in massive high-redshift galaxies with $\mathrm{M_* > 10^{10}~M_\odot}$ at $0.7<z<1.5$ via stacking \citep{Nelson2016b} and in an M31-sized progenitor at $z\sim1.25$ with $\mathrm{M_* \sim 10^{11}~M_\odot}$ \citep{Nelson2019}. 
{These results are} also consistent with {studies of the \ha\ morphologies, which generally find that} the effective radii {of galaxies measured} in \ha\ are larger than that of stellar continuum for star-forming galaxies at $z\sim$ 0.5 -- 3 \citep{Wilman2020, Matharu2022}. 
In addition, this is in line with studies where they found a reduced specific SFR in the centers of massive galaxies at $z \sim1 $ \citep{Nelson2016b}, $z \sim2 $ \citep{Tacchella2018} and at $z \sim 4$ \citep{Jung2017, Ji2023}.

 In section \ref{sec:gradients}, we found that of the 16  galaxies with positive EW(\ha) gradients, {10 also} show positive sSFR gradients. 
The median EW(\ha) gradient is {more positive (0.09 dex kpc$^{-1}$) than both the median sSFR gradient (0.003 dex kpc$^{-1}$) and the median age gradient ($-$0.04 dex kpc$^{-1}$, see Figure~\ref{fig:compare_gradient})}. 
 It is interesting then, that it is not universally true that galaxies with positive EW(\ha) gradients also have positive gradients in sSFR nor negative gradients in age (see Section~\ref{sec:ew}).  

There are multiple reasons we may expect the gradient in  EW(\ha) and the sSFR to behave differently.  
Firstly, {this can be a result of the complex star-formation history within galaxies. Because t}he \ha\ {emission} responds to changes in the SFR on shorter physical timescales compared to the SFR derived from the SED fitting to the broadband data. 
We discuss the implication of this on the relation between the EW(\ha) and sSFR gradients in the context of understanding the spatially-resolved star formation in Section~\ref{sec:discussion_SF}. 
{While the choice of the parameterization of the star-formation history is important, as we argue below this is unlikely to change our overall interpretation (Section~\ref{sec:discussion_SF}).}

{Secondly, other effects could contribute to the gradient in EW(\ha) and the sSFR.  There could be spatial variations in the dust attenuation and/or metallicity that impact our measurements.   We consider this in Section~\ref{sec:dust} and argue that these factors are less likely to be driving the differences between EW(\ha) and sSFR that we observe.}  
%
%
%
%

{Finally, there are exceptions. Two interesting galaxies in our sample show \textit{negative} EW(\ha) and sSFR gradients, and are outliers compared to the rest of our sample.}  {Understanding how these connect to the rest of the sample is important, and we} discuss them in Section~\ref{sec:negative_EW}. 

\subsection{{Spatially Resolved Star Formation Histories in} Galaxies: Evidence of Bulge Formation} \label{sec:discussion_SF}

Comparing the {spatially resolved EW(\ha) and sSFR gradients in galaxies} can provide insights into different timescales in the recent star-formation history (SFH). For example, the \ha-to-UV ratio has been used to quantitatively evaluate SFH burstiness (e.g., \citealp{Guo2016, Faisst2019, Emami2019, Asada2023}).  {Here we can use these results to study the spatially resolved SFHs in galaxies.}

The \ha\ {emission} is sensitive to ionization from O-type stars with lifetimes $\sim$5-10 Myr. The SFR derived from SED fitting primarily responds to changes in the rest-UV continuum, which is sensitive to changes on $\sim$100~Myr timescales (the main-sequence lifetime of B-type stars).  The mass-weighted ages from the SED fitting probe the ages on longer timescales and are therefore sensitive to stars with lifetimes of 100--1000 Myr. Thus, as shown in Figure~\ref{fig:compare_gradient}, the relation between the EW(\ha) and sSFR gradients shows a tighter correlation compared to that of the EW(\ha) and age gradients, where the latter also shows larger scatter.

\begin{figure*}
    \centering
    \includegraphics[width=\textwidth]{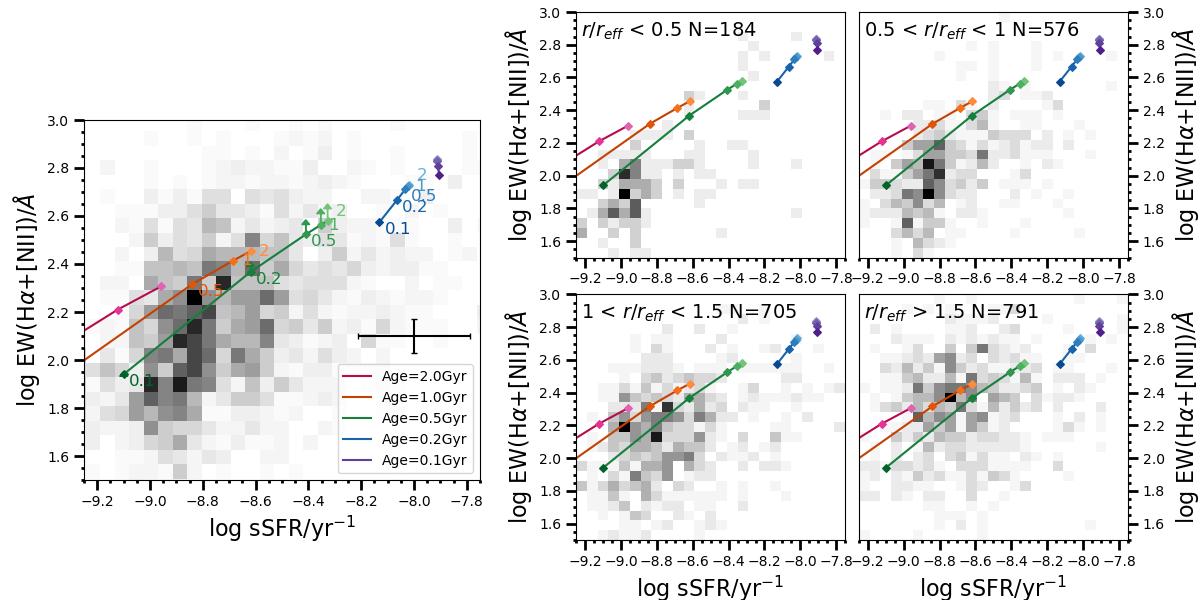}
    \caption{\textit{Left panel:} The grids of sSFR versus EW(\ha) from a set of delayed SFH models overlaid on the 2D histograms of sSFR versus EW(\ha) for all pixels. The {diamonds} show expectations from SED models with delayed-$\tau$ star-formation histories at different ages and with different $\tau$ values. The colors of these crosses correspond to models of different ages and $\tau$ values. The models with the same age are linked with {colored, solid line}. The $\tau$ values are labeled by each model $\tau$=0.1, 0.2, 5, 1, 2~Gyr. The 2D histogram is the same as Figure~\ref{fig:bins} but in grey.  {All models assume Solar metallicity.   The arrows on models at an age = 0.5~Gyr show the shift in the models if we adopt a gas-phase metallicity $Z/Z_\odot = 0.6$ and a stellar metallicity $Z/Z_\odot = 0.16$ (see Section \ref{sec:meta}). The black error bar shows the median error of individual, spatially resolved pixels.} \textit{Right panels}: the 2D histograms for pixels in four bins of galactocentric radius, normalized by the galaxies' effective radii ($r_{\rm eff}$) measured at rest-frame 1.2 $\mu$m (see text). The radii of each bin are listed in each panel. The {colored lines show the same models} as in the left panel.}
    \label{fig:bins_model}
\end{figure*}


Considering that EW(\ha) and the SED-derived sSFR trace the star formation on different timescales, we expect these differences should {manifest} in the position of galaxies in the sSFR -- EW(\ha) {diagram}.
To study this, we used the \bagpipes\ code \citep{Carnall2018} to generate SEDs from stellar population models with simple, delayed, exponentially declining star-formation histories, {where $\mathrm{SFR} \sim t\times \exp(-t/\tau)$, where we allow both the age ($t$) and the $e$-folding timescale ($\tau$)} to vary between 0.1--2 Gyr.  {We also assumed} solar metallicity, a fixed nebular emission with log($U$) = $-$2 and a fixed \citet{Calzetti2000} dust attenuation law with $A(V)$ = 0.4 mag. 
We then computed the EW(\ha) and sSFR from these models.  {The left panel of Figure~\ref{fig:bins_model} overplots these models} on the 2D histogram of sSFR versus EW(\ha) for all pixels from all the galaxies in our sample. 
%
%
Inspection of the figure shows that these models with ages of 0.5 to 2~Gyr and $\tau$ values of 0.1 to 2~Gyr broadly span the parameter space seen in the distribution of EW(\ha) and sSFR within galaxies in our sample.  Therefore, variations in the star-formation histories can account for the distribution of EW(\ha) and sSFR {we observe}. 

We further explored the effects of the star-formation histories on the EW(\ha) and sSFR values as a function of galactocentric radii. 
We separated the galaxy pixels based on their normalized galactocentric radii. Here, we normalized the galactocentric radius of each galaxy by the galaxy's effective radius ($r_\mathrm{eff}$) measured in the bandpass closest to rest-frame 1.2 $\mu$m.\footnote{The imaging bands that corresponds to rest-frame 1.2 $\mu$m are \jwst\ NIRISS F200W for galaxies at $0.6<z\leq1.0$, NIRCam F277W for galaxies at $1.0<z\leq1.6$, and NIRCam F356W for galaxies at $1.6<z\leq2.2$.}  The effective radii are measured using \galfit\ \citep{Peng2002, Peng2010a} with a single S\'{e}rsic profile, applied to the image for each galaxy. We then divided the pixels from each galaxy into four bins of normalized radial distance, $r / r_\mathrm{eff} < 0.5$, $0.5 < r/r_\mathrm{eff} < 1$, $1 < r/r_\mathrm{eff} < 1.5$, and $r/r_\mathrm{eff} > 1.5$, and plot their distribution in sSFR versus EW(\ha). The right four panels of Figure~\ref{fig:bins_model} show these results. 

In these panels of Figure~\ref{fig:bins_model}, we see that as the galactocentric distance increases, the 2D distributions of EW(\ha)--sSFR shift vertically upwards toward higher values of EW(\ha), albeit with large scatter. 
{The median EW(\ha) increases by 0.24 dex from the innermost regions ($r/r_\mathrm{eff} \leq 1$) to the outermost regions ($r/r_\mathrm{eff} > 1$). }
%
Comparing to the \bagpipes\ models, the innermost regions ($r/r_\mathrm{eff} \leq 0.5$) prefer models with relatively young ages ($\sim$500 Myr) but short $e$-folding times, ($\tau$ $\sim$100 Myr).  In contrast, the outermost regions ($1 < r/r_\mathrm{eff} \leq 1.5$ and $r/r_\mathrm{eff} > 1.5$) favor models with large $\tau$ ($\sim$2 Gyr) and a wide range of age ($\sim$0.5-2 Gyr).   
Therefore, if the differences in EW(\ha) and sSFR trace differences in the star-formation history, then these results imply that the innermost regions of the galaxies have experienced a faster build-up of stellar mass (with a star-formation history represented by a short $\tau$, i.e., a starburst), followed by the rapid decay of quenching. This is evident by the ratio of the age/$\tau \sim 5$, which implies the SFRs in the central regions of these galaxies have declined by a factor of $\exp(-\mathrm{age}/\tau) \sim \exp(-5) \approx 1/150$.  Meanwhile, the outer regions of galaxies follow a more steady, slowly varying star-formation history represented by a longer $\tau$ and prolonged ages, where the SFR has declined from its peak by \textit{at most} $\exp(-\mathrm{age}/\tau) \sim \exp(-2/2) \sim 1/3$. 

We note that the values of EW(\ha) and sSFR from the \bagpipes\ models do not change with dust attenuation (i.e., changing $A(V)$), {nor does changing the metallicity have a strong impact on our results. W}e explore the effects of dust {and metallicity} in the next subsection.  

{We also considered how changes to the parameterization of the SFH could impact our interpretation.  Our results depend on the strength of the \ha\ emission and the SED-derived sSFRs, where (as argued above) these are sensitive to changes to the SFH when averaged over $\sim$10~Myr and $\sim$100~Myr timescales, respectively.  The delayed--$\tau$ models, while simplistic, allow for such changes in the relative SFR (i.e., models with short $\tau$ values act as bursts or sudden quenching).  The impact of more complex SFHs would be minimal when averaged on these timescales.  Indeed, \citet{Cutler2023} reconstructed the spatially resolved star-formation histories of the center and outskirts of 66 galaxies at $z\sim2.3$ using the ``flexible'' star-formation histories with \textsc{PROSPECTOR} \citep{Johnson2021}, combining information from \hst\, Spitzer/IRAC, ground-based photometry, and constraints on the gas-phase metallicity from MOSDEF spectroscopy. }
 {They found that for the majority of their galaxies, the outskirts experienced smoother star-formation histories, while the central regions experienced more recent ($<$100 Myr) bursts.  This is similar to our findings and implies our interpretation is qualitatively robust against variations in the assumption of the parameterization of the SFH.  }

This scenario {where the galaxy centroids form in shorter-duration star-formation episodes} can further account for {observations of galaxy sizes measured from spatially resolved imaging} in the mid-IR.  {The mid-IR probes emission from polycyclic aromatic hydrocarbons (PAHs), a tracer of star formation from heating in photo-dissociation regions, (e.g., \citealt{Ronayne2023}).} 
%
%
Early studies with \jwst/MIRI \citep{Magnelli2023, Shen2023} show that the effective radii of galaxies at $0.2 < z < 2.5$ over a range of stellar mass measured in the mid-IR emission are more compact (with smaller effective radii) than either the light in the rest-frame optical/near-IR (which trace stellar light){, the near-UV light (which traces the direct continuum from young stars with $\sim$100 Myr lifetimes), or the \ha\ emission.} 
One explanation for this is that the mid-IR emission tracks star formation over longer timescales than either the UV or H$\alpha$ emission, as longer-lived stars (with lifetimes of 500 Myr -- 1 Gyr) can heat PAHs efficiently \citep{Salim2009, Kennicutt2012, Salim2020}. In this scenario, the mid-IR emission in the central regions of galaxies may be a remnant of the previous starburst episodes, whereas the \ha\ emission, which traces only the most massive, shortest-lived stars, declines faster.  This implies that the galaxy centroids were actively forming in the past $\lesssim 1$~Gyr, and that the SFR has declined since then. If this is correct, then the galaxies in our sample should show similar mid-IR emission concentrated in their centers. This can be tested with studies with \jwst/MIRI in the future. 

 In the outer regions of galaxies, star formation appears to be more steady-state and slowly varying to account for the distribution of EW(\ha)--sSFR seen in our galaxies. 
 Again, this is in line with the ``inside--out'' formation of galaxy disks, and has been inferred from past studies of spatially resolved \ha\ sizes versus the sizes of galaxies in their stellar light at the redshifts of our sample \citep[e.g.,][]{Tacchella2015, Nelson2016b, Wilman2020, Matharu2022}. 
Our results support this scenario, with the added detail that this star formation acts over longer timescales than that responsible for the formation of galaxy centroids {(see also \citealt{Cutler2023})}. 
%


Moreover, our finding is reminiscent of galaxy bulge formation \citep[e.g,][]{Zoccali2003, Ballero2007}. 
The majority of galaxies in our sample have positive EW(\ha) gradients and positive/flat sSFR gradients (Figure~\ref{fig:compare_gradient}), implying that we may be witnessing the end of the formation of galaxy bulges in the galaxies in our sample.  
Theoretical work suggests that the formation of dense stellar cores (e.g., bulges) can stabilize gaseous disks, leading to a local suppression of star-formation \citep[e.g.,][]{Bournaud2007, Martig2009, Genel2012, Sales2012, Dekel2014, Tacchella2019, Semenov2023}. The center regions of galaxies may experience a higher number of discreet starbursts as a result of multiple minor mergers, counter-rotating streams, recycled gas, or tidal compression, which preferentially cause gas to flow toward the galaxy center, triggering star-formation \citep[e.g.,][]{Zolotov2015, Tacchella2016a, Tacchella2016b}. The build-up of these galactic cores may ultimately lead to the formation of the bulge and also to a decrease or cessation of star formation. 
Indeed, the distribution of metallicity in the Milky Way stars shows evidence for such a formation scenario that bulge/thick disk formed through multiple discreet bursts then quench rapidly, where there is more gradual/smooth star-formation in the thin disk \citep[e.g.,][]{Zoccali2003, Ballero2007, Queiroz2021, Queiroz2023}. Therefore, discrete bursts of star formation would account for the EW(\ha)--sSFR and EW(\ha)--age distributions in the innermost galaxy regions in our sample, and possibly indicate bulge formation.

\subsection{{Systematic Effects of Dust and Metallicity}} \label{sec:dust}

{In this section, we discuss other possible effects from dust attenuation, metallicity, and their radial distributions on the measurement of EW(\ha), sSFR, and age and their gradients. }

\subsubsection{{Dust Attenuation}} \label{sec:dust2}

{There are two ways that systematics associated with dust attenuation could impact our interpretation.   The first is that there is a well-known degeneracy between the stellar population age and dust attenuation in SED fitting. Second, as mentioned in the introduction, while the EW(\ha) is generally not affected by dust, there may exist higher attenuation from the birth cloud on the nebular emission, compared to the stellar continuum.  We consider both of these effects here.} 

{The degeneracy} between age and dust attenuation in the SED fitting {is most problematic when} only rest-frame UV-optical data are available (e.g., \citealp{Papovich2001, Conroy2013}).  That is, the same photometry data can be fitted with either a younger model and more extinction or an older model with less extinction. 
However, {we mitigate against this degeneracy as we use imaging from \hst\ and \jwst\ that covers $0.4$ to $\sim$5~$\mu$m. For the galaxies in our sample, this corresponds to coverage through the rest-frame $K$-band, where the effects of dust and stellar population ages are separable (e.g., see next paragraph).  Furthermore, } we found no strong correlation between dust attenuation and mass-weighted age from the results of spatially resolved SED fitting, indicating the SED fitting results are largely not affected by such degeneracy. 

{As an additional check, the} age -- dust degeneracy can be broken when including mid-/far-IR data.  
To test this, we repeated the SED fitting {using} the integrated photometry of the galaxies in our sample including longer-wavelength data in the mid-IR and far-IR {from IRAC/Spitzer at 5.8$\mu$m and 8 $\mu$m, MIPS/Spitzer at 24 $\mu$m, and Herschel/PACS at} 70, 100, 160 $\mu$m.  We cross-matched our catalog with the GOODS-Herschel catalog \citep{Elbaz2011}. We identify 9 galaxies in our sample with $>3\sigma$ MIPS 24$\mu$m detections. For these 9 galaxies, we reran \cigale\ SED fitting using the same parameters as in section \ref{sec:sed}, but now including the additional data. 
We find that when including these mid-IR and far-IR data, the sSFR values decrease slightly, with a median change of 0.07 dex. The mass-weighted ages increase slightly, with a median of 0.03 dex. 
These changes are within the error budget derived on the same quantities without the mid-IR and far-IR data, and some of these changes can be attributed to the associated uncertainties on the mid-IR and far-IR data. 
%

Because we use these same data to derive the spatially resolved SED fits, we expect that our values for the resolved sSFR and ages are not seriously impacted by uncertainties in dust attenuation. 
{In addition, the spatially resolved dust attenuation profiles are smooth for the majority of our galaxies. }
{To further test the effect of dust attenuation, we refit the SEDs for each Voronoi bin for each galaxy while fixing the dust attenuation law to be the same as that from the fit to the galaxy's integrated SED. On average, this increases the sSFR gradients by a median value of 0.08 dex~kpc$^{-1}$ and decreases the age gradients by a median value of 0.06 dex~kpc$^{-1}$.  These are within our error budget and the correlation between the EW(\ha) and sSFR gradients remains significant with a $p$-value of 0.03. }
{Therefore, within the limitations of our dataset, our results are relatively robust to dust attenuation. We expect future analyses including spatially resolved observations of the Balmer decrement to correct for EW(\ha) and/or modeling that includes \jwst/MIRI can probe fully the impact of dust attenuation on our results.}

{One problematic scenario would be if the \ha\ emitting regions experience significantly higher dust attenuation than the continuum.  This would decrease the \ha\ emission more, lowering the EW(\ha).}
{ There are also some }theoretical predictions that dust could directly affect Lyman--\textit{continuum} photons, which would also possibly lead to a reduction of \ha\ \citep{Tacchella2022}.  
%
%
{Previous studies find an increase in dust attenuation toward the center of galaxies at $1 \lesssim z \lesssim 3 $ (e.g., \citealp{Nelson2016a, Wang2017, Tacchella2018, Matharu2023}), but this does not necessarily affect EW(\ha) unless the dust law itself is also changing, or if there exists a deeply obscured, optically thick component. In fact, \citet{Hemmati2015} studied the attenuation ratio between nebular and stellar in eight galaxies at $z\sim0.4$ and found a flat radial profile of the attenuation ratio in the majority of their galaxies. Although this result is based on a small sample, future studies of resolved hydrogen recombination lines (e.g., \ha/\hb) in a large sample would allow constraining the radial profile of attenuation ratio between nebular and stellar and the dust-corrected EW(\ha) radial profile. } 
{Furthermore, for the cases which our galaxies have mid-IR and far-IR detections, we observe only minor differences in the resulting dust attenuation and sSFR values from model fits to the integrated SEDs.  Therefore, it is unlikely that these galaxies have dominant, obscured components missed by the SED modeling that would impact our interpretation.  Nevertheless, future studies of hydrogen recombination lines at longer wavelengths (e.g., Paschen lines) would allow for studies of dust attenuation affecting the nebular regions directly.}

\subsubsection{{Metallicity}} \label{sec:meta}

{Another possible concern is if there are strong metallicity gradients in our galaxies that alter the interpretation of the EW(\ha) and sSFR gradients.  However, we expect this to be a minor effect.   The gas-phase metallicity gradients of galaxies at $0.6 < z < 2.6$ are found to be flat for galaxies with $\mathrm{M_* \gtrsim 10^{10} M_\odot}$ and slightly positive metallicity gradients ($\sim$0.05 dex/kpc) for galaxies with lower masses (e.g., \citealt{Simons2021})}  
{The gradients of gas-phase metallicity could affect the \nii/\ha\ ratio across galaxies and affect the EW(\ha+\nii) gradient. As mentioned in Section \ref{sec:ew}, we measure the EW(\ha+\nii) as EW(\ha) and assume a constant \nii/\ha/ across one galaxy. In the case of a positive metallicity gradient of 0.05 dex~kpc$^{-1}$, the log(\nii/\ha) would increase by 0.09 dex~kpc$^{-1}$ following \citet{Pettini2004}, which would lead to the EW(\ha+\nii) gradient increase by 0.02 dex~kpc$^{-1}$. This change is within the error budget we measure on the EW(\ha) gradients. }

{In addition, we tested the effects of changing the metallicity in the models, but this had only a minor effect on the results.  In Figure~\ref{fig:bins_model} we show the shift in the expected EW(\ha) and sSFRs when using models with a gas-phase metallicity of $Z/Z_\odot = 0.6$ and a stellar metallicity of $Z/Z_\odot = 0.16$ as arrows for models with age = 0.5~Gyr (compared to the canonical Solar metallicity values).  These alternative values for the metallicities are adopted based on measurements from galaxies with similar stellar mass and redshift (\citealt{Sanders2021} and \citealt{Chartab2023}). }
%
{The modeled EW(\ha) values generally increase by a median of $\sim0.06$ dex and increase more with higher EW(\ha) in the range of $-$0.05 to 0.12. These are smaller by more than a factor of two than the median difference (0.24 dex) between the innermost region ($r/r_\mathrm{eff} \leq 1$) and the outermost region ($r/r_\mathrm{eff} > 1$). Thus, although the metallicity cannot fully explain the shift in the EW(\ha) from the galaxy's center to the outskirt region, we cannot rule out the possible effect of the metallicities causing some differences in the sSFR--EW(\ha) histogram, at least partially. }
{We expect future analysis on details of metallicity profiles of these galaxies and using more extended grids of stellar evolutionary models such as Mesa Isochrones and Stellar Tracks \citep[MIST,][]{Dotter2016, Choi2016} could help in understanding the effect of metallicity. }

\subsection{Galaxies with negative EW(\ha) gradients: the duty cycle of Bulge Formation} \label{sec:negative_EW}

Interestingly, two galaxies in our sample show a negative EW(\ha) gradient, a negative sSFR gradient, with a flat age gradient. 
These two galaxies (NGDEEP\_01948 and NGDEEP\_01831) show irregular morphological signatures, possibly indicative of a merger or that they have starbursting clumps near their centers (see Fig. \ref{fig:stamps}). 

To test for mergers, we measured the Gini-M$_{20}$ values for our galaxies using \statmorph\ \citep{Rodriguez-Gomez2019} from the \jwst\ imaging corresponding to rest-frame 1.2~$\mu$m (as above). Compared to the separation from \citet{Lotz2008}, most of our galaxies (16 out of 19) fall well within the region associated with galaxy disks, including one of the irregular galaxies here (NGDEEP\_01948).   Three of the galaxies fall in the region of the Gini-M20 plot associated with mergers, including the other irregular galaxy here (NGDEEP\_01831), but they are all very close to the merger-disk separation line. Therefore, it is inconclusive if these two irregular galaxies are mergers.  

Inspecting the spatially resolved sSFR and age maps of these two galaxies, NGDEEP\_01948 and NGDEEP\_01831 (see Figure~\ref{fig:maps_for3}), they both have centers with high sSFR and low ages, suggesting rapid growth. Therefore we favor the scenario where these galaxies have recently undergone a nuclear starburst, which could result from a recent gas flow, or as a massive star-forming clump(s), which may form and migrate from the disk \citep{Dekel2009b, Dekel2009a, Zolotov2015, Tacchella2016b}.  Other galaxies at these redshifts indicate possible central gas accretion events based on ``inverted'' metallicity gradients, which could be taken as evidence for metal-poor cold-gas inflows from the ISM or circumgalactic medium depositing directly onto the centers of galaxies (e.g., \citealp{Wang2019, Simons2021}). 
The latter are rare (occurring in less than 10\% of the samples), which is consistent with the two galaxies we find with central regions with elevated sSFRs and high EW(\ha).

Therefore, these two galaxies may be the ``exceptions that prove the rule'' in our interpretation of the EW(\ha), sSFR, and age gradients.  
If we return to the argument {from Section~\ref{sec:discussion_SF} that bulge formation is episodic, with a short duty cycle of star formation, then we may have caught these two galaxies in a growth phase of their central regions}. 
Taking the ratio of the number of galaxies with forming centroids (two) to the total (19) within the period of cosmic time (4.9~Gyr) spanned by the redshift range of our sample ($0.6 < z < 2.2$), this duty cycle is approximately $\approx (2/19) \approx 10$\%, and the timescale for this period is 10\% $\times 4.9$~Gyr = 500 Myr.  This is remarkably similar to the ages of the galaxy cores in our sample, supporting our argument that we are witnessing bulge formation in this sample.

Taken together, our data imply that most (16 out of 19) bright, star-forming galaxies at $0.6 < z < 2.2$ form stars in disks that grow in an ``inside-out'' fashion.  At the same time, they are forming their bulges through multiple, short-duration bursts.   In some cases (two out of 19 in our sample) there are indications that we have caught galaxies during a formation stage, likely as a result of a recent increase in the nuclear gas-accretion rate or via clump migration.

\section{Summary} \label{sec:summary}

We study the \ha\ equivalent width maps of 19 galaxies at $0.6 < z < 2.2$ from NIRISS slitless spectroscopy as part of the NGDEEP survey. Our sample is dominated by bright galaxies that lie along the star-formation main sequence with a stellar mass range of $\mathrm{10^9 - 10^{11} M_\odot}$. We combine these data with ancillary deep \hst\ and \jwst\ imaging from CANDELS, JADES, and JEMS, from which we perform spatially-resolved SED fitting with 21-band fluxes for these galaxies and construct their specific SFR and mass-weighted age maps.  The resolved SED fits and \ha\ maps resolve angular structures in galaxies on scales of $0\farcs18$, allowing quantitative measurements of nebular emission lines and galaxy properties at a spatial resolution of $\sim$1 (proper) kpc.  

The EW(\ha) and sSFR are independent measurements, derived from the NIRISS grism data and multi-band SED fitting, respectively. They both trace the current star formation activity, and the mass-weighted ages from the SED fitting probe the time since the last major star formation episode. 
We compare the spatially-resolved EW(\ha), sSFR, and mass-weighted age and their radial profiles. Our results are summarized as follows:  

\begin{itemize}
    \item We find strong correlations between EW(\ha) and sSFR, and an anticorrelation between EW(\ha)  and the mass-weighted age. This is true for both the integrated emission from galaxies and on a  pixel-by-pixel scale.  Therefore, spatially within galaxies, the EW(\ha) increases with increasing sSFR and the EW(\ha) increases with decreasing age.  
    
    \item Quantifying the radial profiles of EW(\ha), sSFR and mass-weighted age, most galaxies show positive EW(\ha), positive sSFR gradients, and negative age gradients. The median values of the gradients in EW(\ha), sSFR, and age gradients are {$0.09^{+0.03}_{-0.06}$} dex~kpc$^{-1}$, {$0.003^{+0.049}_{-0.038}$} dex~kpc$^{-1}$ and {$-0.04^{+0.04}_{-0.03}$} dex~kpc$^{-1}$. 
    This indicates that, for most galaxies, their outer regions exhibit higher levels of star formation and appear to be younger when compared to their central regions.  This is consistent with the inside-out formation of galaxies. 

    \item We compare our measured EW(\ha) and sSFR to stellar population obtained from a set of star formation histories following a delayed-$\tau$ model using \bagpipes. We find that these models roughly span the observed distribution of EW(\ha) and sSFR and EW(\ha) and age.    
    We then compare the distributions of EW(\ha) and sSFR to the models as a function of galactocentric radius (normalized by the galaxy effective radii). The central regions of galaxies ($r/r_\mathrm{eff} \leq 0.5$) favor models with a small age ($\sim$500 Myr) and rapidly declining star-formation histories with $\tau \sim$100 Myr.  In contrast, the outer regions ($r > r_\mathrm{eff}$) favor models with slowly evolving star-formation histories with larger $\tau \sim$2 Gyr, and a wide range of age ($\sim$0.5-2 Gyr).  We argue that this indicates that while most galaxies grow their disks in an ``inside-out'' fashion, their cores (i.e., bulges) grow through multiple, short-duration bursts. {However, there are important systematics that could impact this result, including assumptions about the dust law, metallicity gradients, and parameterization of the star-formation history.  We discuss these uncertainties and offer predictions that our results make that can be tested in future studies (in Section~\ref{sec:discussion}). }
    %

    \item Two galaxies show negative EW(\ha) and sSFR gradients with a flat {(i.e., constant)} age gradient. These two galaxies show irregular morphologies, but we do not find strong evidence of a merger from their morphologies (using Gini-M20).  We, therefore, conclude these galaxies are experiencing central star-forming episodes, either as a result of recent gas accretion in their cores or through the migration of clumps from their disks. Comparing the ratio of the number of galaxies in our sample with active central star formation (two) to the total (19), gives a short duty cycle of 10\%, albeit with a large {statistical} uncertainty.  This suggests that bulge formation {may} occur at these redshifts in multiple star-formation episodes of $<500$~Myr.  Based on the age of Galactic bulge stars and the redshift of our sample, it is {possible} that we are witnessing the end of the period of star formation in the galaxies in our sample. 

\end{itemize}

This paper demonstrates the capability of \jwst/NIRISS slitless spectroscopy data to construct the nebular emission line maps of high-redshift galaxies, which can be combined with broad-band imaging to gain useful constraints on the spatially resolved formation of galaxies during the peak of the cosmic SFR density.  In the future, we will expand on this work by adding the second half of the NGDEEP NIRISS observations, which will provide a large sample of high-redshift galaxies with secure nebular emission lines (\ha, \hb, \oii, \oiii) maps for star formation, dust attenuation, and metallicity studies, which can inform a more complete picture of galaxies' growth and quenching.   We also expect to expand this work using larger samples of galaxies with broad-band imaging, extending to the mid-IR (with \jwst\ MIRI) and WFSS from NIRISS.

\begin{acknowledgments}
We acknowledge the hard work of our colleagues in the NGDEEP collaboration and everyone involved in the \jwst\ mission.   
This work benefited from support from the George P. and Cynthia Woods Mitchell Institute for Fundamental Physics and Astronomy at Texas A\&M University. 
CP thanks Marsha and Ralph Schilling for generous support of this research. 
This work is based on observations made with the NASA/ESA/CSA {\it JWST}. The data were obtained from the Mikulski Archive for Space Telescopes at the Space Telescope Science Institute, which is operated by the Association of Universities for Research in Astronomy, Inc., under NASA contract NAS 5-03127 for \jwst. These observations are associated with program \#2079. 
Some/all the {\it JWST} data presented in this paper were obtained from the Mikulski Archive for Space Telescopes (MAST) at the Space Telescope Science Institute. The specific observations analyzed can be accessed via 
\cite{https://doi.org/10.17909/3s7h-8k54}, \cite{https://doi.org/10.17909/fsc4-dt61} and \cite{https://doi.org/10.17909/8tdj-8n28}. 
 JM is grateful to the Cosmic Dawn Center for the DAWN Fellowship. 
PGP-G acknowledges support from grant PGC2018-093499-B-I00 and PID2022-139567NB-I00 funded by Spanish Ministerio de Ciencia e Innovación MCIN/AEI/10.13039/501100011033, FEDER, UE. 
The work of CCW is supported by NOIRLab, which is managed by the Association of Universities for Research in Astronomy (AURA) under a cooperative agreement with the National Science Foundation. 

\end{acknowledgments}

%

\vspace{5mm}





\appendix

\section{Resolved Maps for individual Objects} \label{app:maps}

\begin{figure*}
    \centering
    \includegraphics[width=0.45\textwidth]{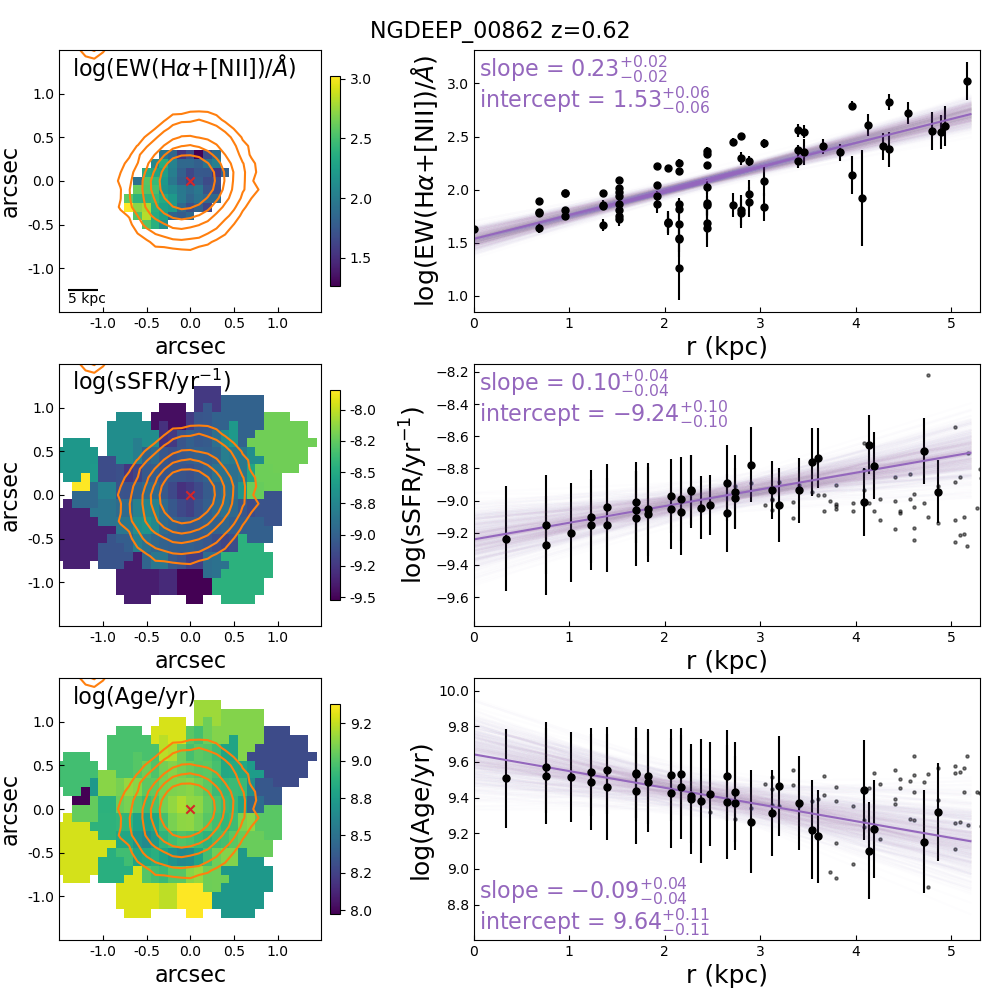}
    \includegraphics[width=0.45\textwidth]{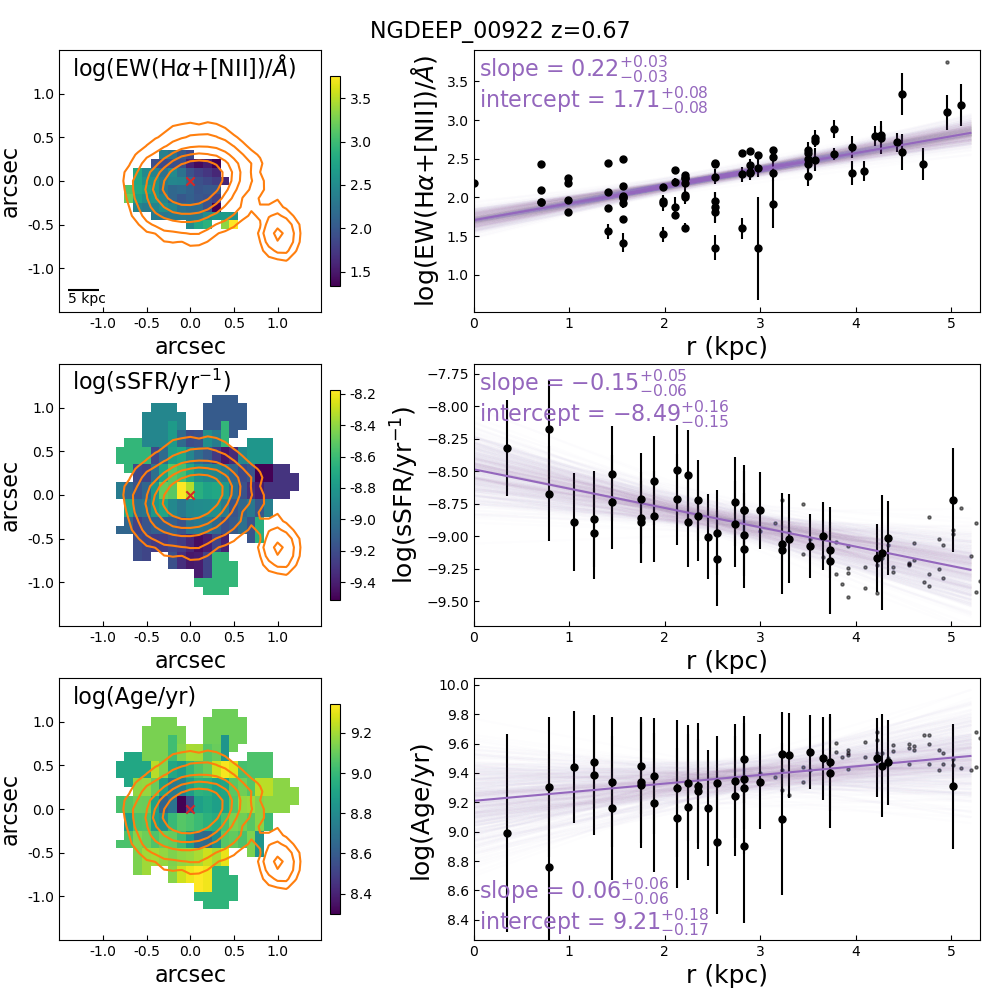}
    \includegraphics[width=0.45\textwidth]{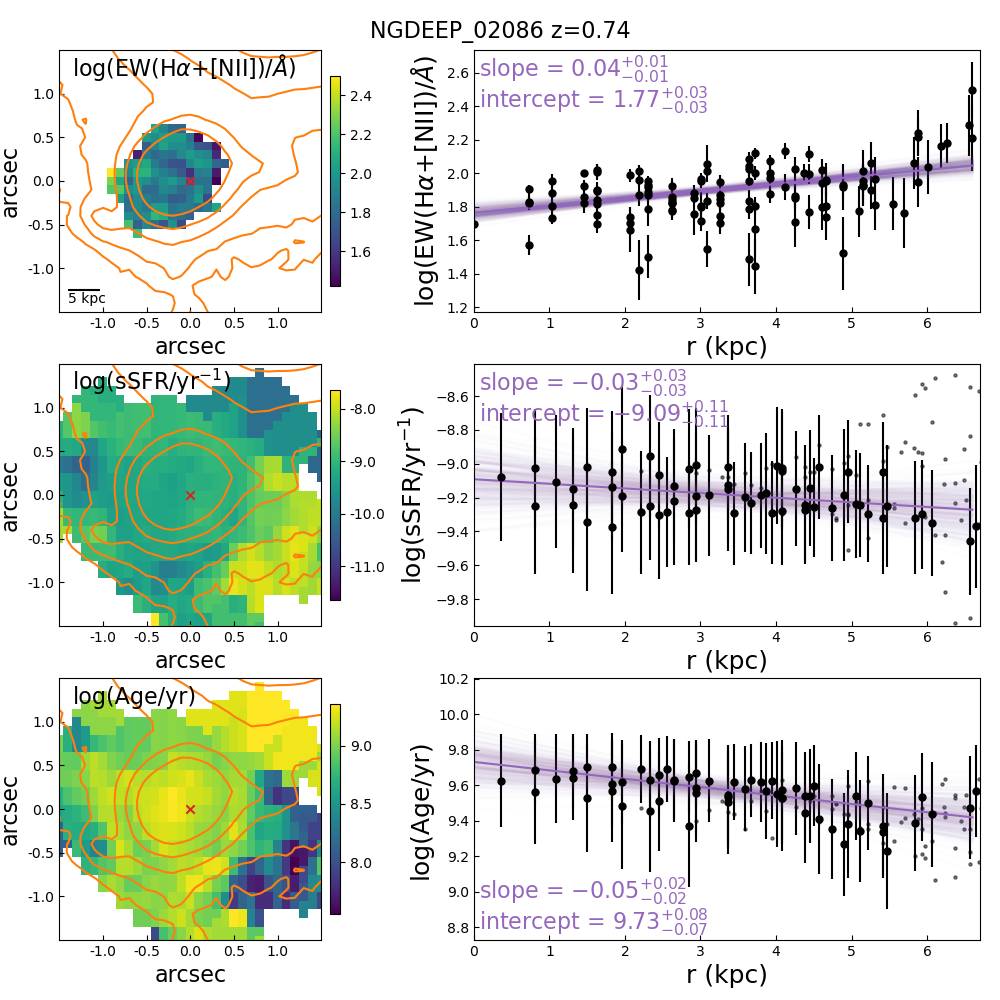}
    \includegraphics[width=0.45\textwidth]{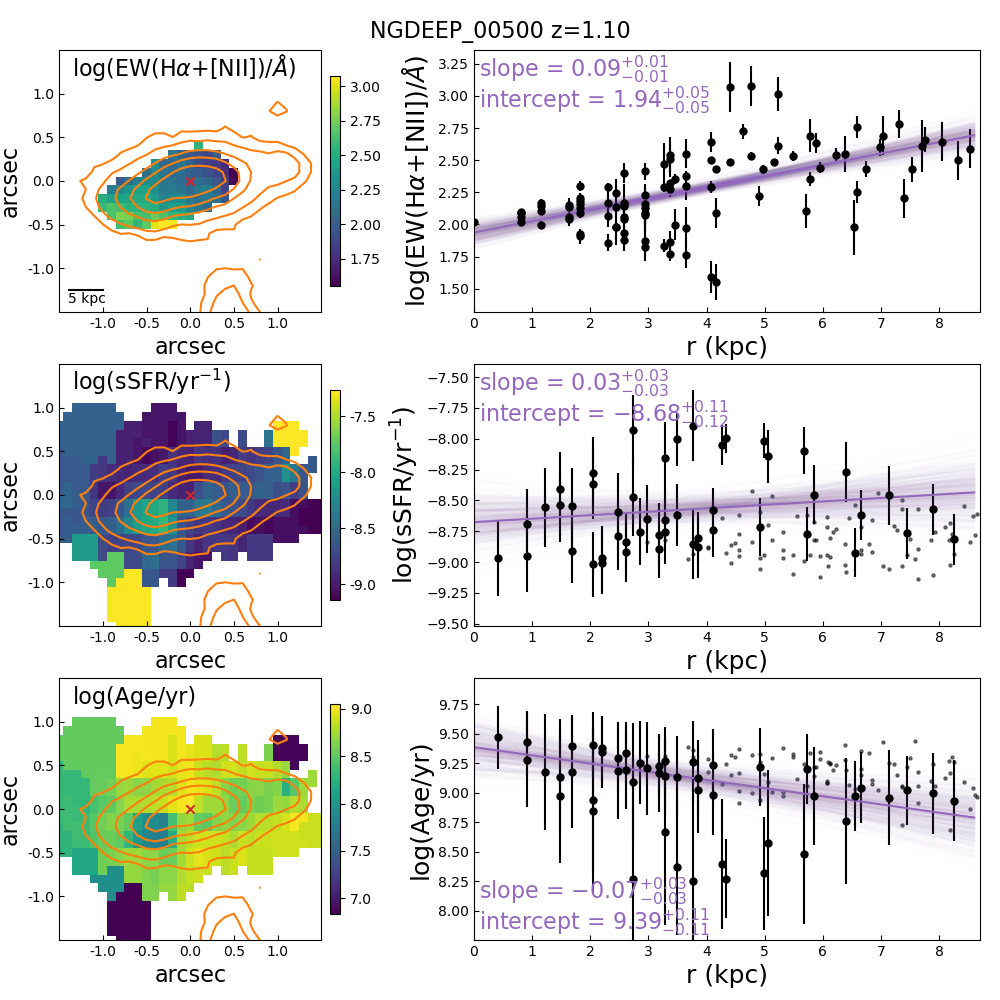}
    \includegraphics[width=0.45\textwidth]{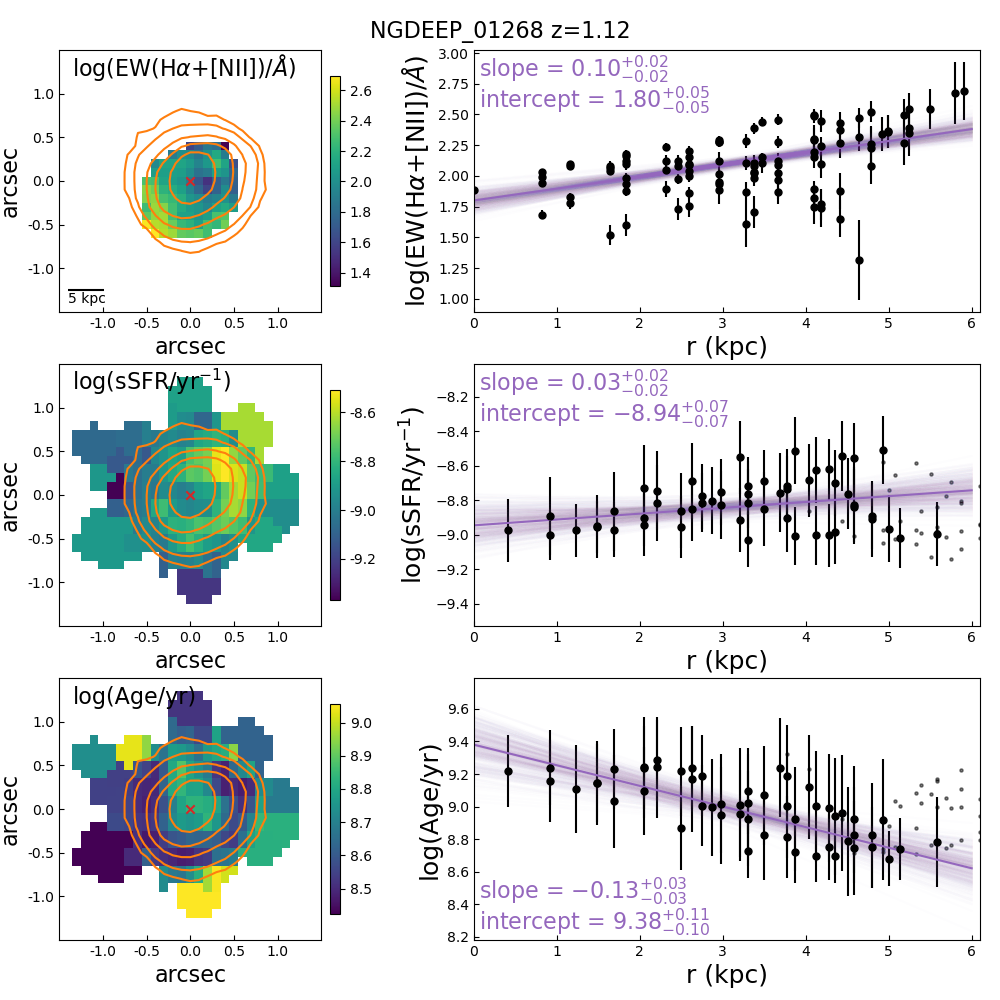}
    \includegraphics[width=0.45\textwidth]{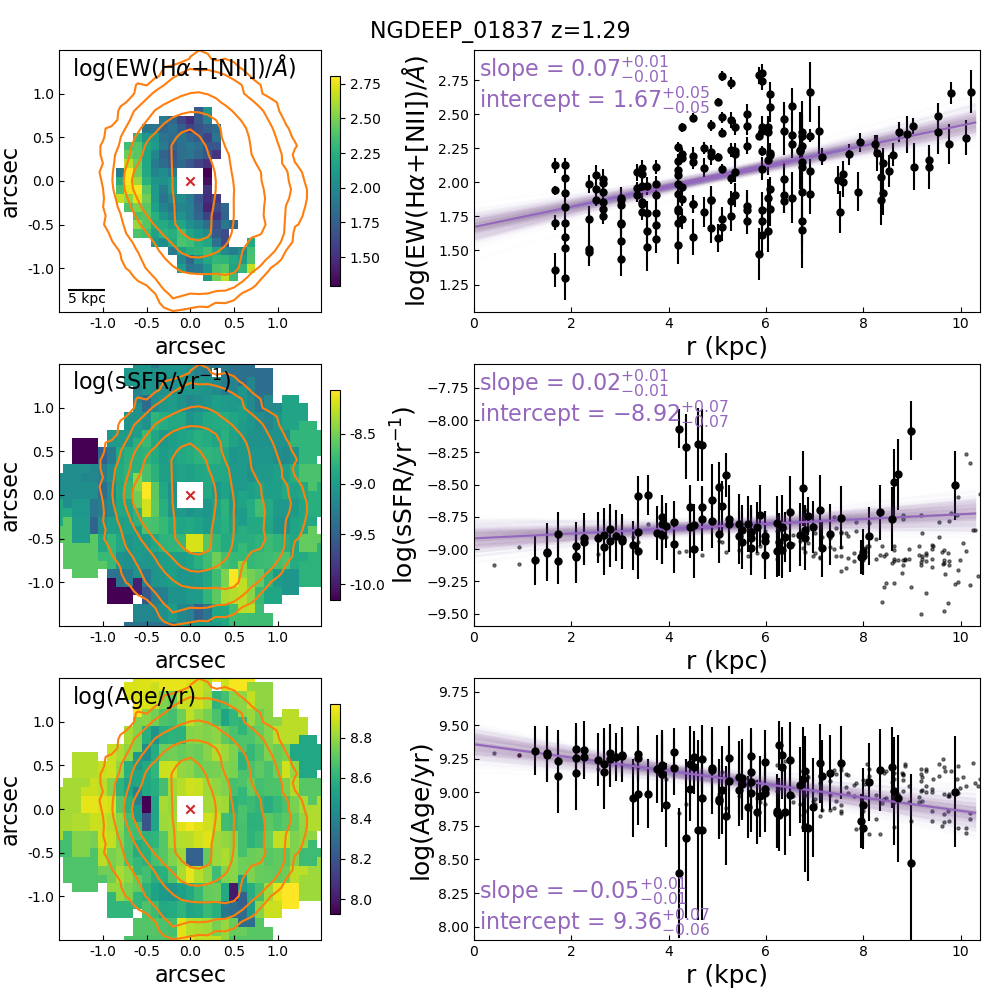}
    \caption{The EW(\ha), sSFR, and age maps and their radial gradients for 6 galaxies in our sample. The same as Figure \ref{fig:radialprofiles} for each galaxy. }
    \label{fig:radialprofiles_app1}
\end{figure*}

\begin{figure*}
    \centering
    \includegraphics[width=0.45\textwidth]{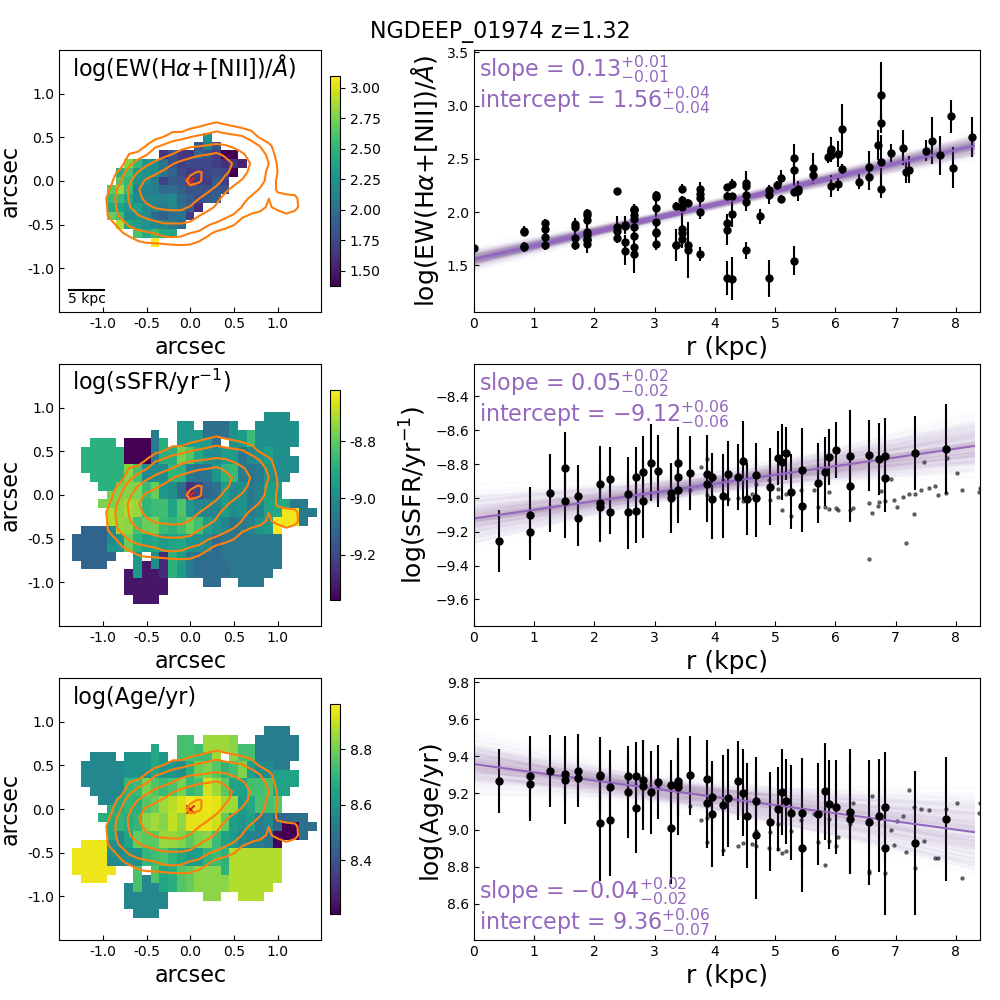}
    \includegraphics[width=0.45\textwidth]{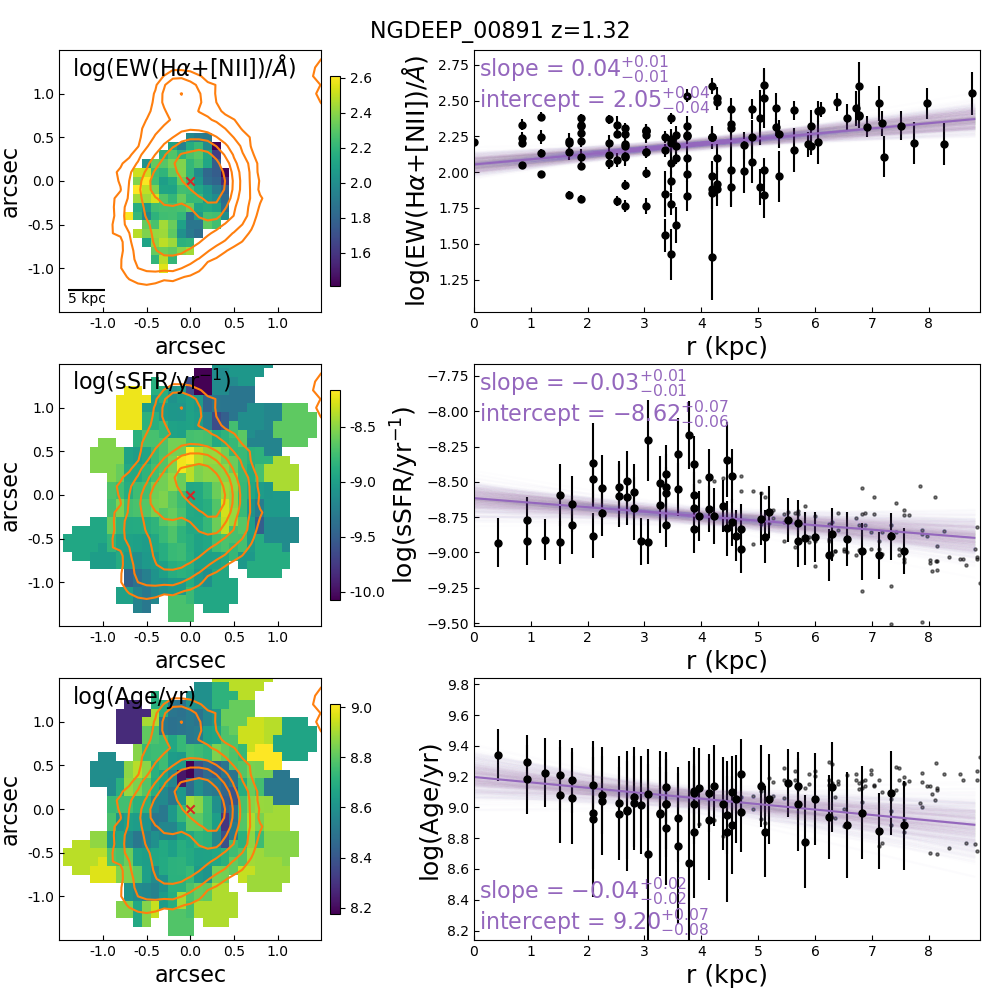}
    \includegraphics[width=0.45\textwidth]{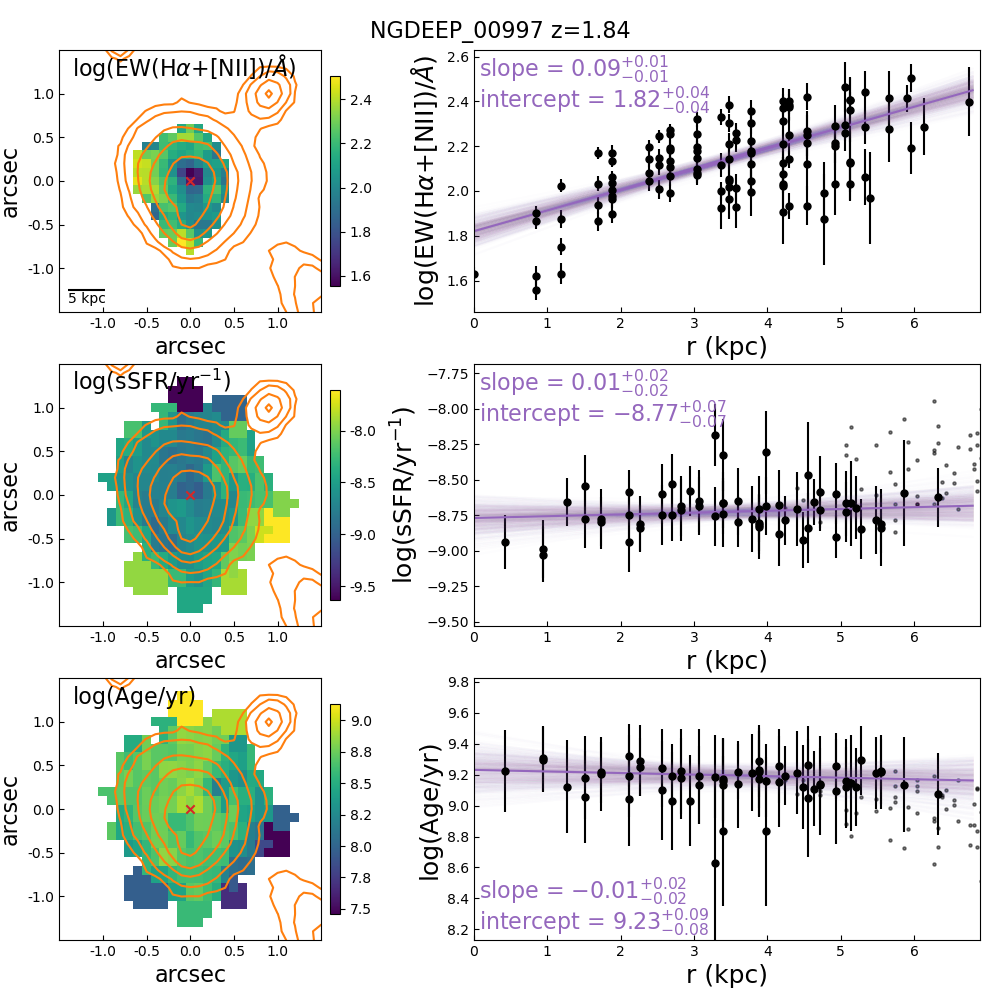}
    \includegraphics[width=0.45\textwidth]{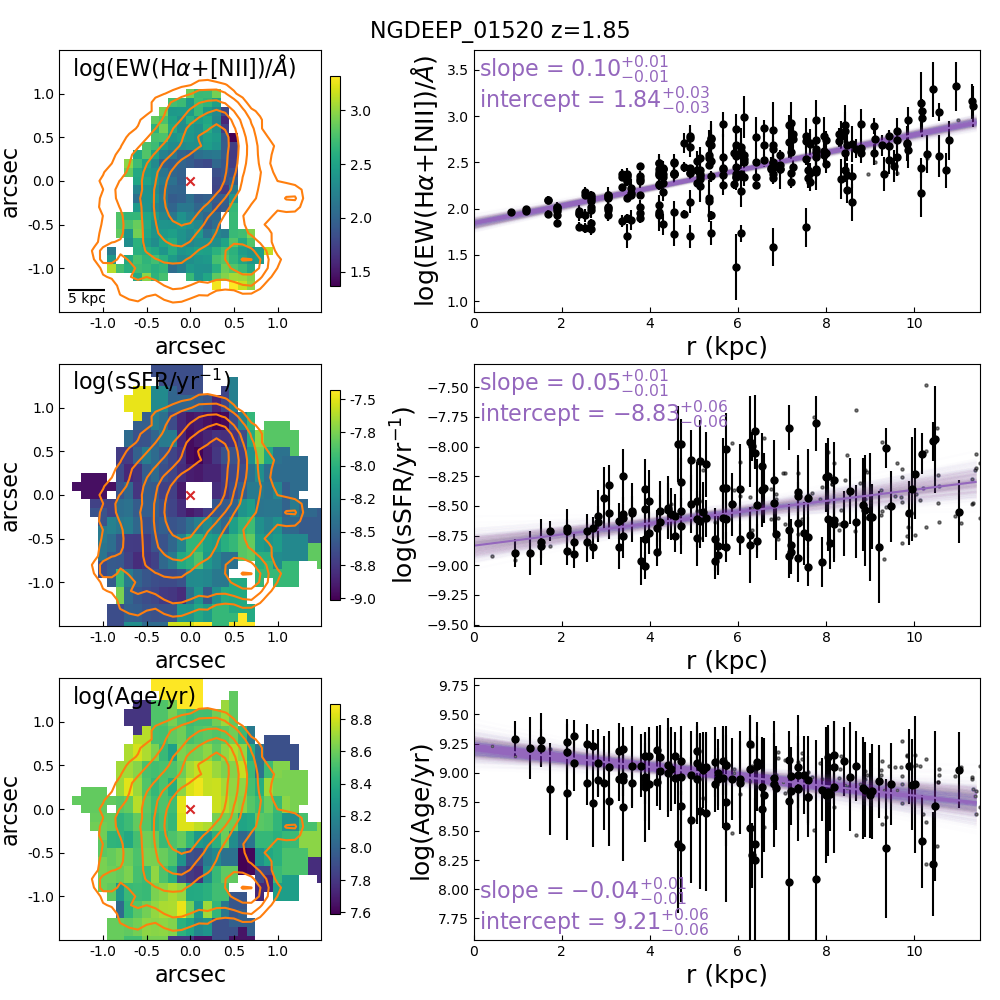}
    \includegraphics[width=0.45\textwidth]
    {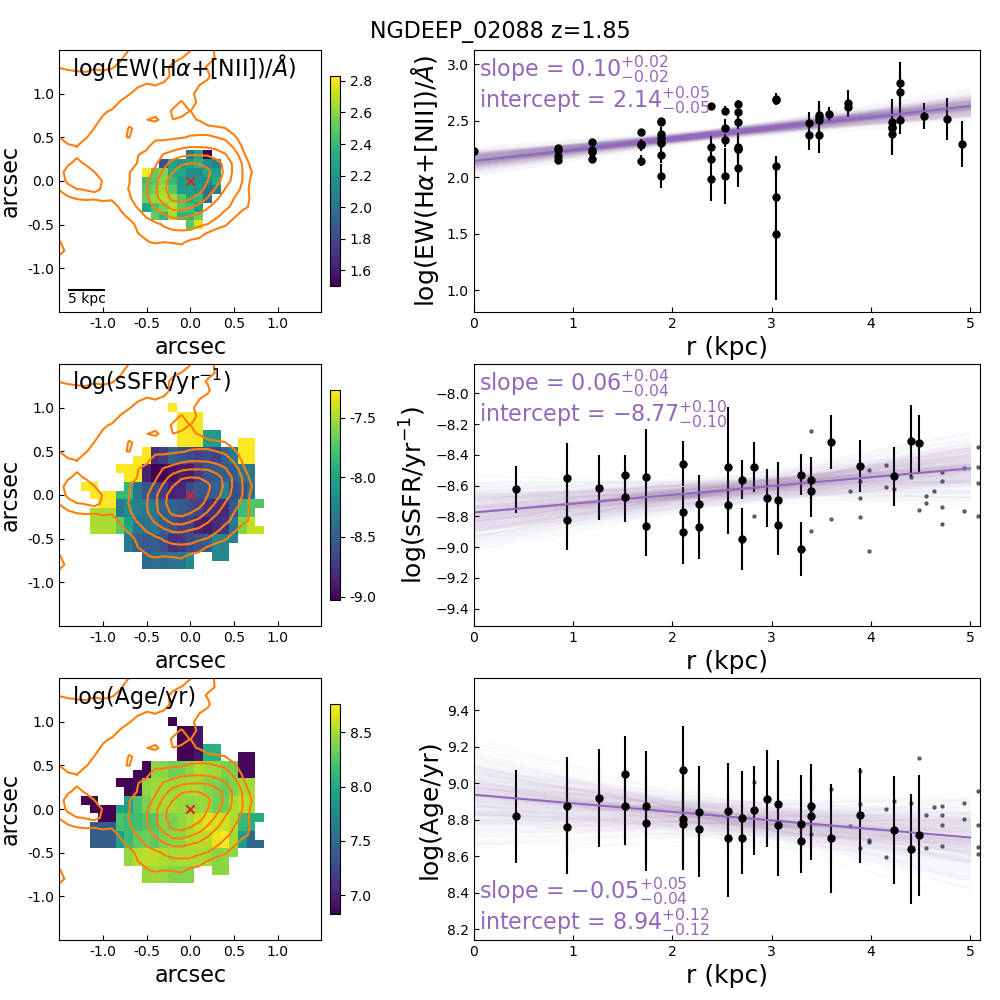}
    \includegraphics[width=0.45\textwidth]{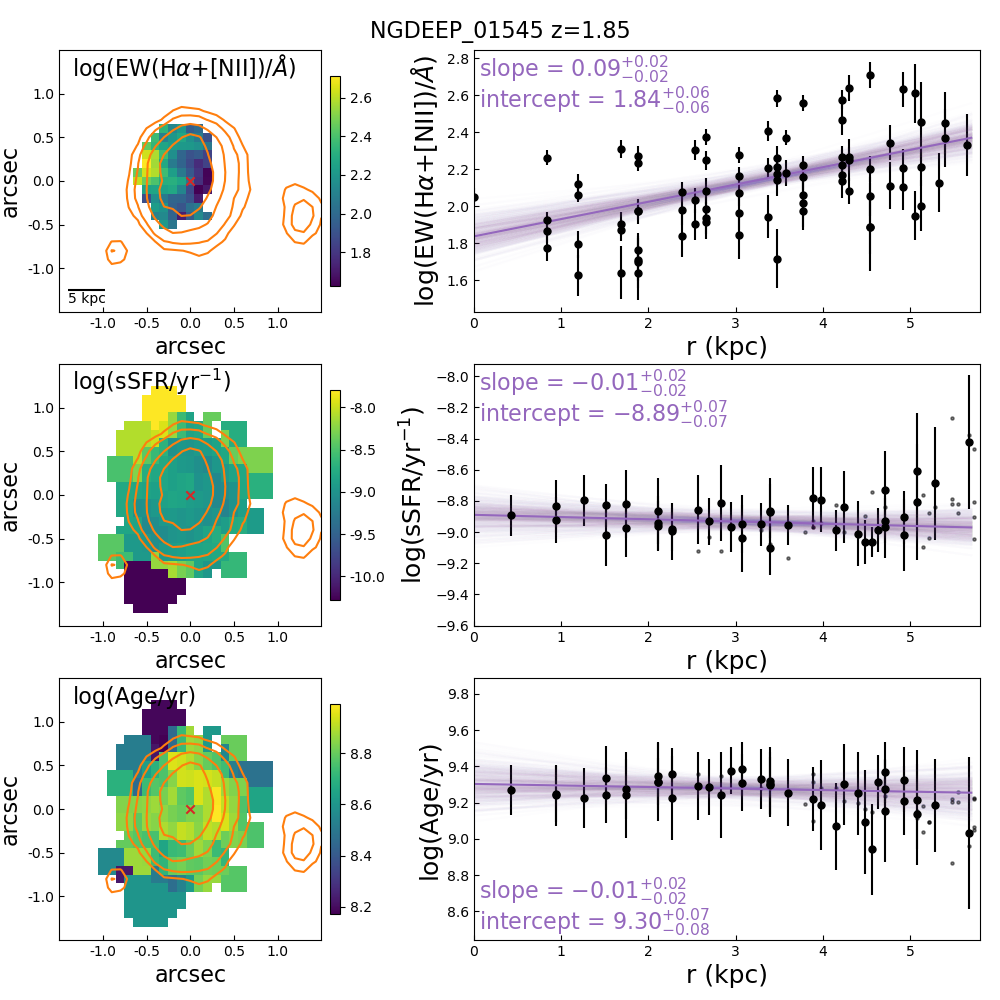}
    \caption{The EW(\ha), sSFR and age maps and their radial gradients for 6 galaxies in our sample. The same as Figure \ref{fig:radialprofiles} for each galaxy.}
    \label{fig:radialprofiles_app2}
\end{figure*}

\begin{figure*}
    \centering
    \includegraphics[width=0.45\textwidth]{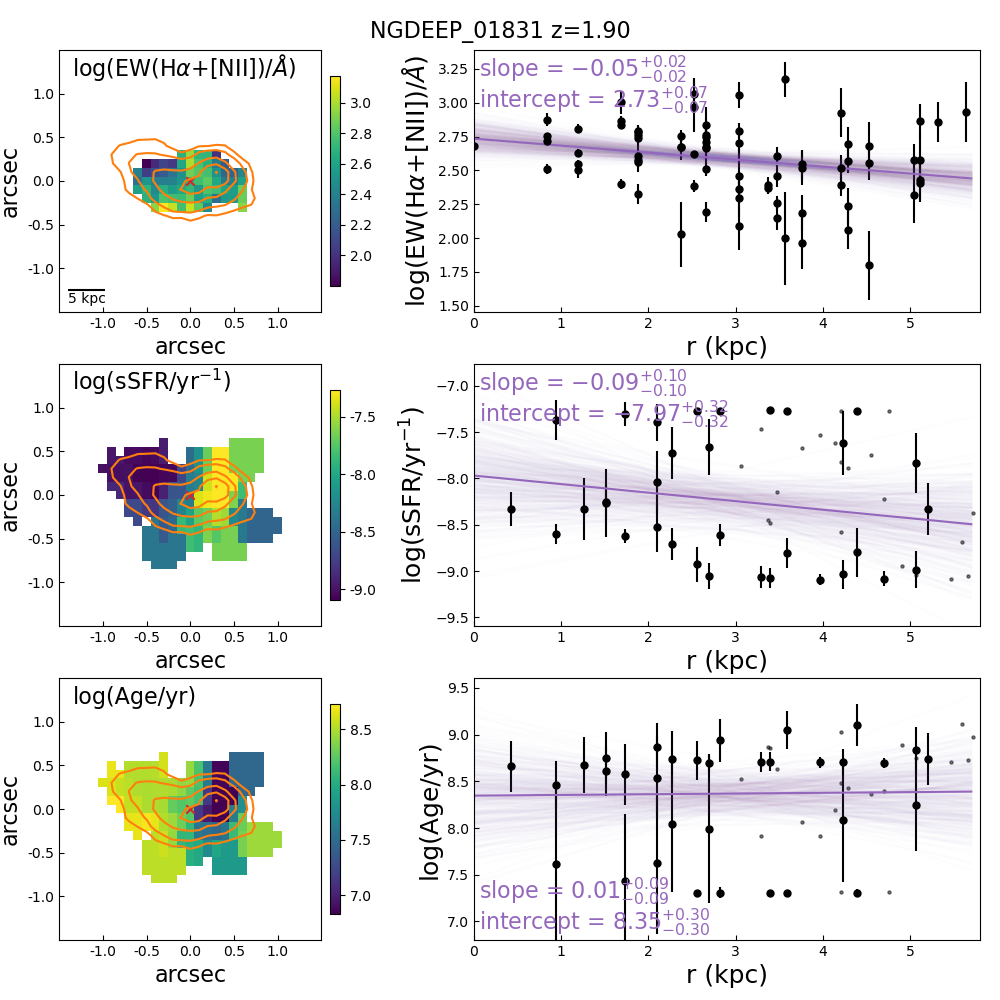}
    \includegraphics[width=0.45\textwidth]{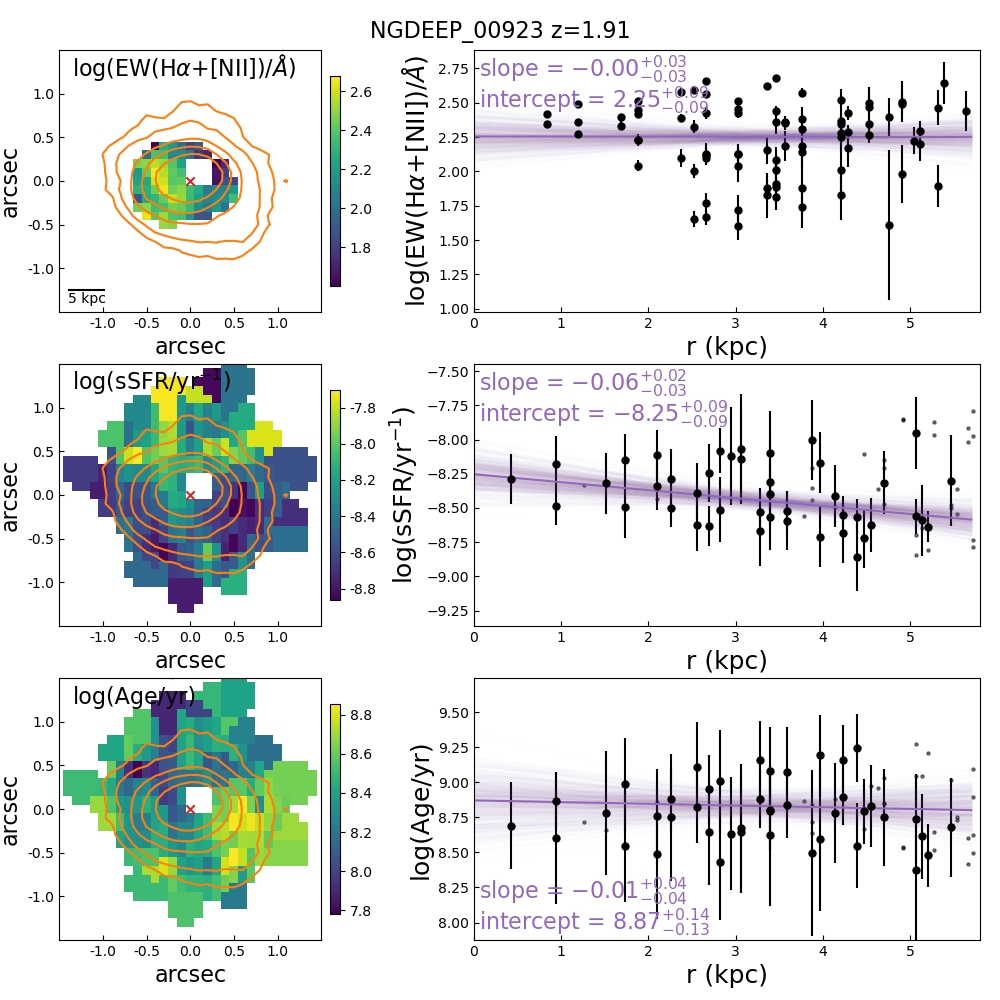}
    \includegraphics[width=0.45\textwidth]{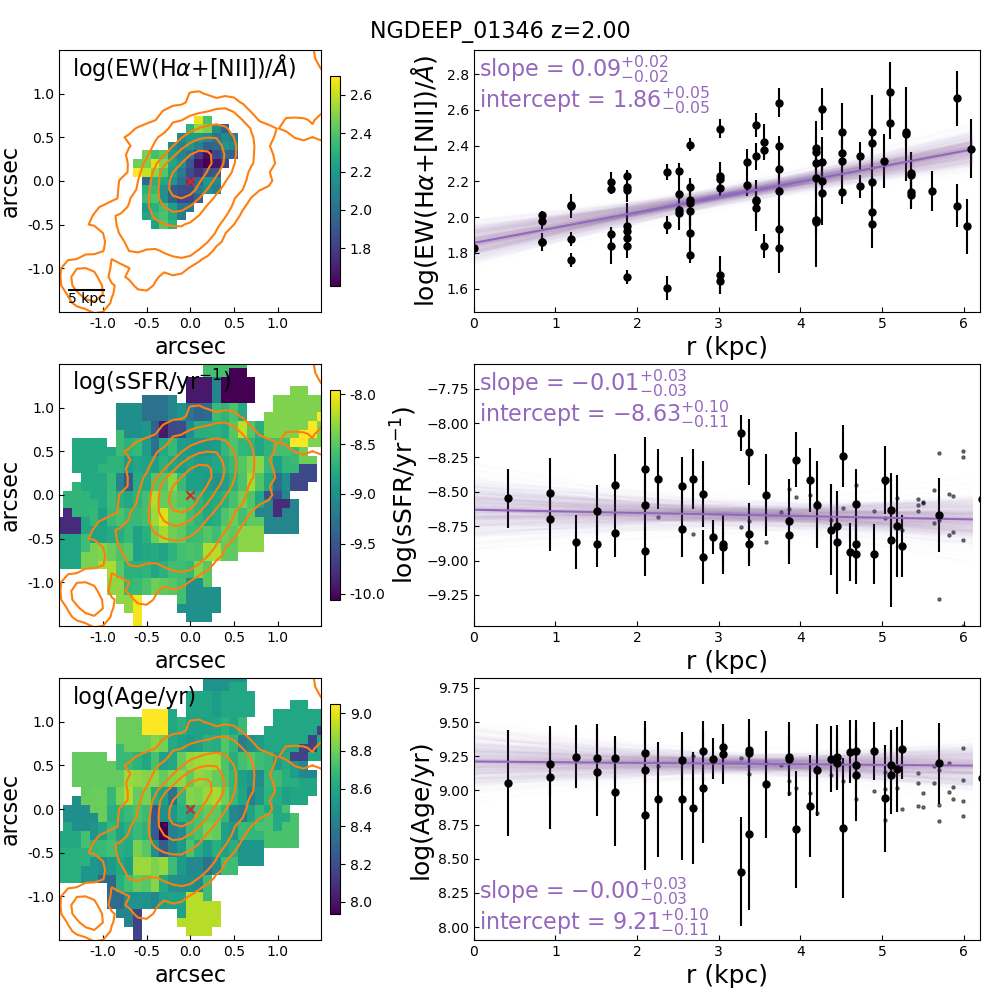}
    \includegraphics[width=0.45\textwidth]{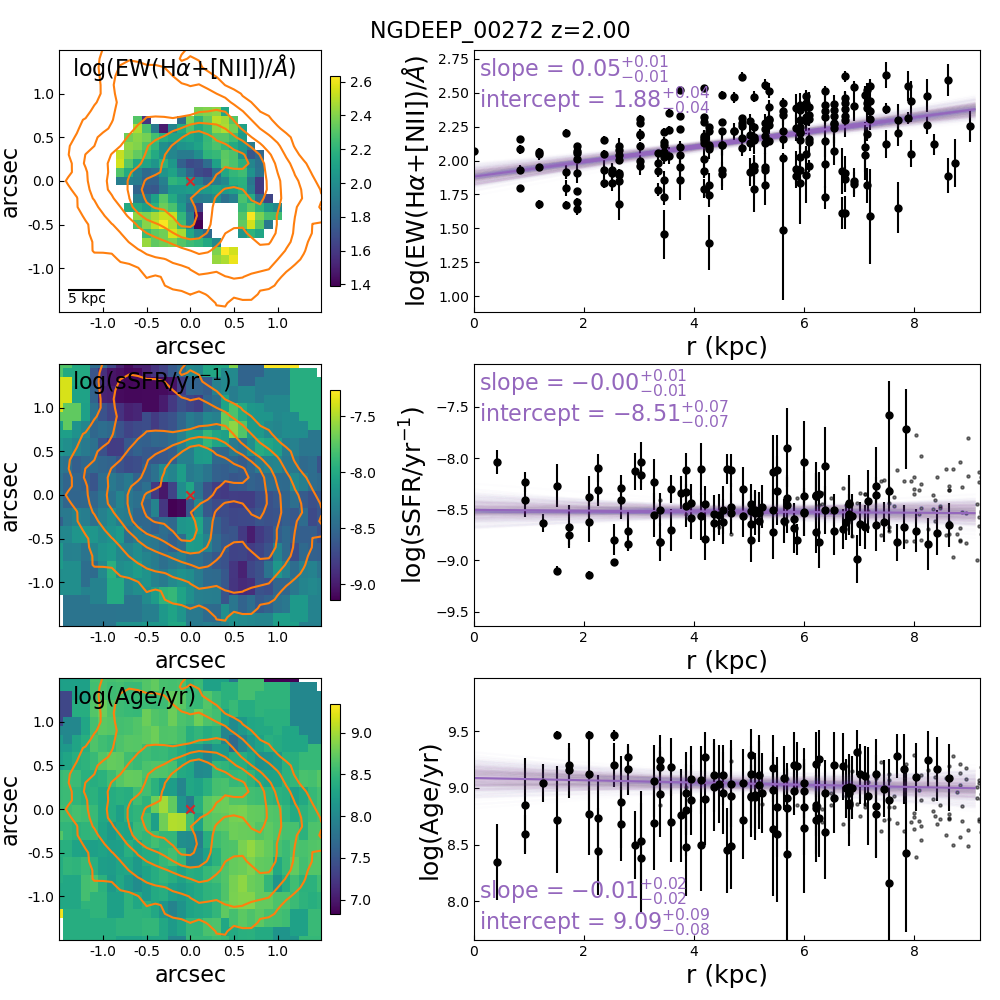}
    \includegraphics[width=0.45\textwidth]
    {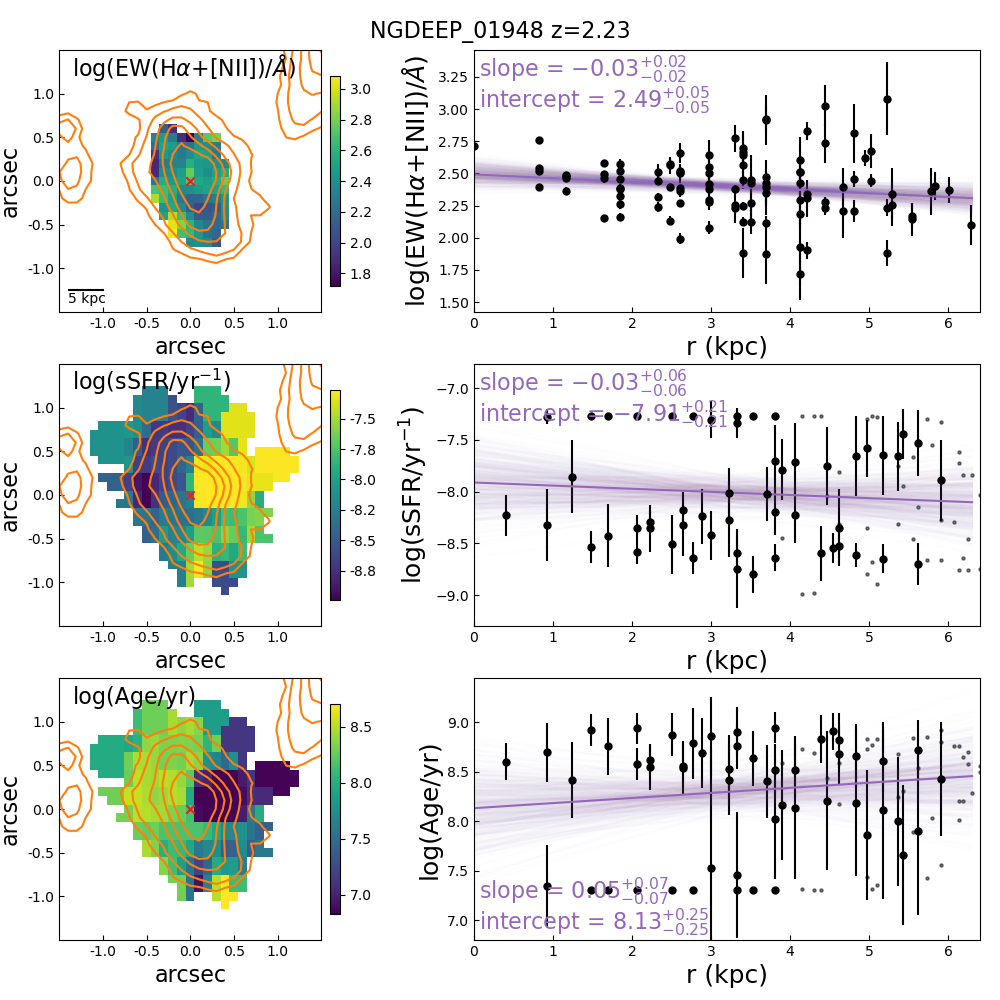}
    \caption{The EW(\ha), sSFR and age maps and their radial gradients for 5 galaxies in our sample. The same as Figure \ref{fig:radialprofiles} for each galaxy.}
    \label{fig:radialprofiles_app3}
\end{figure*}

This Appendix shows Figures~\ref{fig:radialprofiles_app1}, \ref{fig:radialprofiles_app2}, and \ref{fig:radialprofiles_app3}, which show maps of the EW(\ha), sSFR, ages, and their galactocentric gradients for the full sample (except for the galaxies shown in Figure~\ref{fig:radialprofiles} and Figure~\ref{fig:radialprofiles2}). 

\bibliography{ref}{}
\bibliographystyle{aasjournal}



\end{document}